\documentclass[12pt]{article}
\usepackage{amsmath, amsfonts}
\usepackage{siunitx}
\usepackage{graphicx,psfrag,epsf}
\usepackage{enumerate}
\usepackage{natbib}
\usepackage{amsthm}
\usepackage{bm}

\newtheorem{example}{Example}
\newtheorem{definition}{Definition}
\newtheorem{proposition}{Proposition}
\usepackage{url} 
\usepackage[colorlinks,citecolor=blue,urlcolor=blue,filecolor=blue]{hyperref}
\usepackage{multirow}

\newcommand{\blind}{1}

\begin{document}

\def\spacingset#1{\renewcommand{\baselinestretch}%
{#1}\small\normalsize} \spacingset{1}


\if1\blind
{
  \title{\bf  Generalized Bayesian nonparametric clustering framework for high-dimensional spatial omics data}
  
  \author{Bencong Zhu \\ 
    Department of Statistics, The Chinese University of Hong Kong \\
    Guanyu Hu \\
    Department of Biostatistics and Data Science,\\ The University of Texas Health Science Center at Houston\\
    Xiaodan Fan$^{*}$ \\
    Department of Statistics, The Chinese University of Hong Kong  \\
    and \\
    Qiwei Li$^{*}$ \\
    Department of Mathematical Sciences, The University of Texas at Dallas \\
    }
  \maketitle
  \clearpage
} \fi

\if0\blind
{
  \bigskip
  \bigskip
  \bigskip
  \begin{center}
    {\LARGE\bf Generalized Bayesian nonparametric clustering framework for high-dimensional spatial omics data}
\end{center}
  \medskip
} \fi

\bigskip
\begin{abstract}
The advent of next-generation sequencing-based spatially resolved transcriptomics (SRT) techniques has transformed genomic research by enabling high-throughput gene expression profiling while preserving spatial context. Identifying spatial domains within SRT data is a critical task, with numerous computational approaches currently available. However, most existing methods rely on a multi-stage process that involves \textit{ad-hoc} dimension reduction techniques to manage the high dimensionality of SRT data. These low-dimensional embeddings are then subjected to model-based or distance-based clustering methods. Additionally, many approaches depend on arbitrarily specifying the number of clusters (i.e., spatial domains), which can result in information loss and suboptimal downstream analysis. To address these limitations, we propose a novel Bayesian nonparametric mixture of factor analysis (BNPMFA) model, which incorporates a Markov random field-constrained Gibbs-type prior for partitioning high-dimensional spatial omics data. This new prior effectively integrates the spatial constraints inherent in SRT data while simultaneously inferring cluster membership and determining the optimal number of spatial domains. We have established the theoretical identifiability of cluster membership within this framework. The efficacy of our proposed approach is demonstrated through realistic simulations and applications to two SRT datasets. Our results show that the BNPMFA model not only surpasses state-of-the-art methods in clustering accuracy and estimating the number of clusters but also offers novel insights for identifying cellular regions within tissue samples.

\end{abstract}

\noindent%
{\it Keywords:}  High-dimensional data, spatially resolved transcriptomics, factor analysis, Gibbs-type priors, Markov random field
\vfill

\newpage
\spacingset{1.45} 
\section{Introduction}
\label{sec:intro}


Recent advancements in sequencing technology have greatly accelerated the development of both single-cell RNA sequencing (scRNA-seq) and spatially resolved transcriptomics (SRT). Unlike scRNA-seq, SRT not only measures gene expression but also maps the precise locations of thousands of genes within tissue samples. This capability has made SRT an essential tool in uncovering novel insights in biomedical research \citep{marx2021method}. SRT protocols are generally classified into two categories: image-based and next-generation sequencing (NGS)-based \citep{zhang2021spatial}. Among the image-based approaches, techniques such as sequential single-molecule fluorescence \textit{in situ} hybridization (seqFISH) \citep{lubeck2014single}, multiplexed error-robust FISH (MERFISH) \citep{chen2015spatially}, and spatially-resolved transcript amplicon readout mapping (STARmap) \citep{wang2018three} offer high spatial resolution but are limited to mapping hundreds of transcripts and costly setups. In contrast, NGS-based techniques like spatial transcriptomics (ST) \citep{staahl2016visualization}, 10X Visium platform \citep{vickovic2019high}, and Slide-seq \citep{rodriques2019slide} provide a more accessible means to quantify the whole transcriptome on the spot level, albeit with comparatively lower spatial resolution.


Spatial domain identification is one of the key issues in SRT data analysis \citep{yuan2024benchmarking}. It involves defining clinically or biologically meaningful regions by segmenting the whole domain with similar molecular characteristics or morphological features. This crucial step lays the groundwork for numerous subsequent analyses, including spatial domain-based differential expression, trajectory mapping, and functional pathway analysis \citep{thrane2018spatially, moses2022museum}. 
Considering an $p$-by-$n$ numeric matrix $\bm{X}$ representing the raw or processed SRT molecular profile, where $n$ and $p$ denotes the number of cells/spots and genes, respectively, and each entry is $x_{ji}$ for $i=1,\ldots,n$ and $j=1,\ldots,p$. From a statistical perspective, the objective is to cluster cell/spot-wise gene expression profiles $\bm{x}_{i} = (x_{i1}, \ldots, x_{ip})^{\top}$ for $i = 1, \ldots, n$ into mutually exclusive components. Under the model-based clustering framework, each observation is assumed to be drawn from a mixture distribution
\begin{equation}
    \bm{x}_{i}|\cdot \quad\sim\quad \sum_{h=1}^{H}\pi_{h}\mathcal{K}(\bm{x};\theta_{h}), \label{eq:mix_dist}
\end{equation}
where $H$ is the number of component (i.e., spatial domains), $\bm\pi = (\pi_{1}, \ldots, \pi_{H})^{\top}$ are the component weights, and $\mathcal{K}(\bm x; \theta_{h})$ is the kernel density function for component $h$. 

Adapting the above framework to SRT clustering analysis presents three main challenges. First, like other omics data, SRT data, particularly those generated from NGS-based platforms, are high dimensional, with $p\gg n$. However, under the normal assumption $\mathcal{K}(\bm{x}; \theta_{h}) = N(\bm\mu_{h}, \Sigma_{h})$, \cite{chandra2023escaping} proved that as $p \rightarrow \infty$, observations tend to converge into a single cluster. In response, most existing model-based clustering approaches for SRT data, such as BayesSpace \citep{zhao2021spatial}, SCMEB \citep{yang2022sc}, BASS \citep{li2022bass}, and iIMPACT \citep{jiang2024iimpact}, employ a tandem approach. This typically involves first reducing the dimensionality of $\bm{X}$ using techniques like principal component analysis (PCA), followed by fitting a classical mixture model on the low-dimensional embeddings. However, these multi-stage methods have at least two notable drawbacks: 1) the separated optimization of loss functions for dimension reduction and clustering potentially leads to inconsistencies in cluster allocation \citep{markos2019beyond} and 2) the dimension reduction step overlooks the uncertainty in low-dimensional features, treating them as error-free during the subsequent clustering analysis. Some recent methods, such as BNPSpace \citep{zhu2023bayesian}, BayeCafe \citep{li2023interpretable}, and BINRES \citep{yan2024bayesian}, have replaced dimension reduction with feature selection to analyze SRT data. However, these approaches often require computationally intensive searches across all $p$ genes, and their success heavily depends on the accuracy of the feature selection process.

Second, the number of components $H$ is typically unknown. Thus, classical mixture model formulations that rely on a pre-specified $H$ are less appealing due to their inadbility to account for the uncertainty in estimating the number of non-empty clusters \citep{legramanti2022extended}. A promising solution is the use of Bayesian nonparametric approaches, which replace the Dirichlet-multinomial prior with alternatives that allow for adaptive growth in the cluster number $H$ as the sample size $n$ increases. Techniques like the Dirichlet process mixtures (DPM) \citep{ferguson1973bayesian} and mixture-of-finite-mixtures (MFM) \citep{miller2018mixture} facilitate flexible clustering in recent decades. However, many resulting methods \citep[see e.g.,][]{li2017bayesian, fruhwirth2021generalized, hu2023bayesian, zhao2023spatial, zhu2023bayesian, meng2024bayesian} have been developed independently, and a general Bayesian nonparametric model-based clustering framework for high-dimensional spatial omics is lacking. 
To bridge this gap, we propose a framework based on Gibbs-type priors \citep{de2013gibbs, camerlenghi2024contaminated}, which unifies many aforementioned Bayesian nonparametric formulations, standardizing common properties and advancing computational and inferential strategies. 

Last but not least, incorporating spatial information through appropriate priors is essential for enhancing clustering accuracy and biological interpretability. For SRT data from platforms like ST and 10X Visium, where spots are arranged in square and triangle lattices, neighboring spots' labels can be effectively modeled using an MRF or Potts model \citep{li2022bass, liu2022joint, zhao2021spatial, yan2024bayesian}. Recent studies have shown that using the Markov random field (MRF) model significantly improves the accuracy of spatial domain identification \citep{zhu2018identification,zhao2021spatial,jiang2024iimpact,li2023interpretable}. However, integrating spatial information via MRF model in the aforementioned Gibbs-type priors is still unexplored. 

To address the methodological challenges outlined above, we propose a novel class of Bayesian nonparametric mixture of factor analysis (BNPMFA) models. To the best of our knowledge, BNPMFA represents the first unified framework that 1) directly models high-dimensional gene expression data, preserving the integrity of the information, 2) simultaneously estimates the number of spatial domains $H$ and spatial domain configurations with MRF-constrained Gibbs type priors while ensureing the consistent estimation, and 3) rigorously quantifies the uncertainty in spatial domain identification. This comprehensive and scalable approach represents a major breakthrough in the analysis of high-dimensional spatial omics data. The effectiveness of BNPMFA is validated through realistic simulations, and its application to two benchmark SRT datasets significantly enhances our capability to identify and characterize distinct cellular regions within tissue samples, advancing the field of spatial omics data modeling and analysis.

The structure of this paper is organized as follows. Section \ref{sec:model} introduces the proposed BNPMFA, which combines a mixture of factor analysis with MRF-constrained Gibbs-type priors. In Section~\ref{sec:infer}, we present theoretical results on model identifiability and outline the procedure for posterior inference. Sections \ref{sec:sim} and \ref{sec:app} demonstrate the performance of various Gibbs-type prior models through simulation studies and apply the most effective model to two types of SRT data. Finally, Section \ref{sec:con} concludes with a summary and a discussion of potential directions for future research.

\section{Model}
\label{sec:model}

A typical SRT dataset can be summarized into two profiles: the molecular and geospatial profiles. In addition to the molecular profile represented by the aforementioned gene expression matrix $\bm{X}$,  the geospatial profile is depicted by a numeric matrix $\bm{T} \in \mathbb{R}^{n \times 2}$, where the $i$-th row $\boldsymbol{t}_{i\cdot}=(t_{i1},t_{i2})\in\mathbb{R}^2$ gives the x and y-coordinates of the $i$-th spot. The spatial information for spots on a square or triangle lattice allows us to define the geospatial profile by an $n\times n$ binary adjacency matrix $\bm{E}$, where each entry $e_{ii'}=1$ if spot $i$ and $i'$ are neighbors and $e_{ii'}=0$ otherwise. For spots not located on a lattice structure, We can derive $\bm{E}$ from $\bm{T}$ by calculating the Euclidean distances between all spot pairs, i.e., $e_{ii'}={I}(\sqrt{(t_{i1}-t_{i'1})^2+(t_{i2}-t_{i'2})^2}<c_0)$, where ${I}(\cdot)$ denotes the indicator function and $c_0\in\mathbb{R}^+$ is the desired threshold. Note that all diagonal entries in $\bm{E}$ are set to zero by default, i.e., $e_{ii}=0,\forall i$.

\subsection{Model formulation}
Suppose $\{\bm x_{i}\}_{i=1}^{n}$ are drawn from an unknown $p$-dimensional density $f \in \mathcal{F}$, with $\mathcal{F}$ the set of densities with respect to Lebesgue measure on $\mathbb{R}^{p}$ . We model this multivariate density via the following mixture factor analysis model:
\begin{equation}
\begin{aligned}
\bm{x}_{i} &= \bm{W} \bm{y}_{i} + \bm{\epsilon}_{i}, \\
  \bm y_{i} \mid z_{i} = h &\sim \mathcal{K}(\bm{y}; \theta_{h}),\\ \label{eq:model}
 \bm{\epsilon}_i & \sim N(\bm{0}, \Lambda),\\
\end{aligned}
\end{equation}
where $\bm{y}_{i} \in \mathbb{R}^{q}$ with $q < p$, $\bm{W} = [w_{jl}]_{p \times q}$ are factor loadings, and $\mathcal{K}(\bm{y}; \theta_{h})$ is the kernel density for component $h$. Throughout the paper we will assume a diagonal residual covariance $\Lambda = \text{diag}(\sigma_{1}^{2}, \ldots, \sigma^{2}_{p})$. Instead of assuming $y_{i} \sim N(0, \bm{I}_{q})$ in the classic factor analysis model, we assign a mixture distribution on the latent factors to group observations with an unknown number of components. Under the assumption $\mathcal{K}(\bm y; \theta_{h}) = N(\bm{\mu}_{h}, \Sigma)$, the marginal distribution of $\bm{y}$ is mixture normal distribution with $\bm y_{i} \sim \sum_{h\geq 1}\pi_{h}N(y;\bm \mu_{h}, \Sigma)$ where $\pi_{h} = P(z_{i} = h)$. Note that the kernel density $\mathcal{K}(\bm{y};\theta_{h})$ could be any distribution. We assume the covariances across different components are invariant for simplicity and computational efficiency. The latent variable $\bm{y}$ and $z$ can be treated as nuisance latent variables which are marginalized out in joint distribution $P(\bm x, \bm y, z \mid \bm{W}, \{\bm \mu_{h}\}_{h\geq1}, \Sigma, \Lambda )$ by integration, obtaining
\begin{equation}
    f(\bm{x} \mid \bm{W}, \{\bm \mu_{h}\}_{h\geq 1}, \Sigma, \Lambda) = \sum_{h\geq 1} \pi_{h}N(\bm x; \bm{W}\bm\mu_{h}, \bm{W}\Sigma\bm{W}^{\top} + \Lambda).
\end{equation}

The proposed model differs fundamentally from the popular mixture of factor analyzers introduced by \cite{ghahramani1996algorithm}, which defines a mixture of multivariate Gaussians at the observed data level in $p$-dimensional space, where each cluster has cluster-specific means and covariance matrices, with the dimension of the covariances reduced \textit{via} the factor analysis model. In contrast, our approach focuses on learning a shared affine space, enabling us to define a simple location-scale mixture of Gaussians. By enforcing $\text{rank}(\bm{W}\Sigma\bm{W}^{\top}) = q$, we effectively impose a rank constraint on the covariance structure of the observational matrix. A similar framework has been proposed by \cite{baek2009mixtures} with a fixed number of components and solved with an EM algorithm. 

We adopt a conjugate prior to ensure computational efficiency for the prior distribution of the model parameters. Specifically, the factor loadings are assigned normal priors with $w_{jl} \sim N(0, \tau_{w}^{-1}\sigma^{2}_{j})$ for $j \in \{1, \ldots, p\}$ and $l \in \{1, \ldots, q\}$. The variances are assigned with inverse gamma distribution by $\sigma^{2}_{j} \sim \text{Ga}(a, b)$ for $j \in \{1, \ldots, p\}$. For the parameters on latent space, we use Jeffreys prior of normal distribution, which is $\mu_{h} \sim N(0, \tau_{\mu}^{-1}\Sigma)$ with $\Sigma \propto |\Sigma|^{-\frac{q+1}{2}}$ for $h \geq 1$. 

\subsection{Gibbs-type priors}
One natural approach to defining the generative process for the partition variable $\bm{z}$ is to employ a Dirichlet-multinomial prior distribution, obtained by marginalizing the group probability $\bm{\pi} = (\pi_{1}, \ldots, \pi_{H})^{\top} \sim \text{Dir}(\beta, \ldots, \beta)$ from the multinomial likelihood for $\bm{z}$. When the number of groups, $H$, is fixed and finite, this formulation corresponds to the Bayesian mixture of the factor analysis model. However, as mentioned earlier, the number of spatial domains in SRT data is often unknown and needs to be inferred from the data itself. A possible solution is placing a prior on $H$, which leads to the mixture-of-finite-mixtures (MFM) \citep{miller2018mixture}. Another option is a Dirichlet process partition mechanism corresponding to the infinite relational model. Such an infinite mixture model differs from MFM in that $H = \infty$, implying that infinitely many observations would give rise to infinitely many clusters. 

Notably, all the above solutions are specific instances of Gibbs-type priors \citep{de2013gibbs, camerlenghi2024contaminated}, defined on the cluster membership space $\bm z$. For $a > 0$, donate the ascending  factorial with $(a)_{n} = a(a+1)\cdots(a+n-1)$ for any $n \geq 1$ and $(a)_{0} = 1$. The probability function of group $\bm z$ is of Gibbs-type if and only if 
\begin{equation}
    P(\bm{z}) = V_{n}(H) \prod_{h=1}^{H}(1 - \delta)_{n_{h}-1} \label{Gibbs_prior},
\end{equation}
where $n_{h} \in \mathbb{N}$ is the number of spots in domain $h$, $\delta < 1$ denotes the discount parameters, and $\mathcal{W} = \{V_{n}(H): 1 \leq H \leq n\}$ is a collection of non-negative weights. Gibbs-type priors represent a broad and tractable class whose predictive distribution implies that membership indicators $\bm z$ can be obtained in a sequential and interpretable manner \citep{lijoi2007bayesian} according to
\begin{equation}
    P(z_{n+1}=h\mid \boldsymbol{z}) \propto 
\begin{cases}
V_{n+1}(H)(n_{h} - \delta) & \text{for } h = 1, \ldots, H \\
V_{n+1}(H+1) & \text{for } h = H+1 \\
\end{cases}.
\end{equation}

Hence, the group assignment process can be interpreted as a straightforward seating mechanism, where a new spot is assigned to an existing cluster $h$ with probability proportion to the current size reduced by a discount parameter, and further scaled by the weight $V_{n+1}(H)$. Alternatively, the new spot is assigned to a new cluster with probability proportion to $V_{n+1}(H+1)$. This versatile mechanism presents numerous potential generative processes within a unified modeling framework. The examples presented below demonstrate how commonly employed priors can be derived as special cases.

\begin{example}
    (Dirichlet process) Let $\delta = 0$ and set $V_{n}(H) = \frac{\beta^{H}}{(\beta)_{n}}$. Then we can derive DP urn scheme: $P(z_{n+1} = h|\boldsymbol{z}) \propto n_{h}$ and $P(z_{n+1} = H+1|\boldsymbol{z}) \propto \beta$. 
\end{example}
    
\begin{example}
    (Pitman-Yor process) Let $\delta \in [0, 1)$ and set $V_{n}(H) = \prod_{h=1}^{H-1}(\beta + h\delta)/(\beta+1)_{H-1}$ for some $\beta > - \delta$. The we can derive PY urn scheme: $P(z_{n+1} = h|\boldsymbol{z}) \propto n_{h} - \delta$ and $P(z_{n+1} = H+1|\boldsymbol{z}) \propto \beta + H\delta$. 
\end{example} 

\begin{example}
    (Mixture of finite mixtures) Let $\delta < 0$ and set $V_{n}(H) = \sum_{m\geq H} p(m) \beta^{H}/(m\beta + 1)_{n-1}\prod_{h=1}^{H-1}(m-h)$ for $\beta = -\delta$, where $p(\cdot)$ is the assigned prior distribution on the number of clusters. We can derive the MFM urn scheme: $P(z_{n+1} = h|\boldsymbol{z}) \propto V_{n+1}(H)(n_{h} + \beta)$ and $P(z_{n+1} = H+1|\boldsymbol{z}) \propto V_{n+1}(H+1)$ . 
\end{example}

In the MFM model, the prior on the number of clusters ensures the posterior consistency for the estimated number of clusters as $n \rightarrow \infty$ \citep{miller2018mixture}. Instead, the Dirichlet process prior is known to yield inconsistent estimates for the number of clusters when the data is generated from a finite mixture model \citep{miller2013simple}. Therefore, we suggest Gibbs-type priors with $\delta < 0$ in practice to avoid the generation of extraneous clusters, such as a mixture of finite mixtures prior. In this study, we specifically consider the MFM model with $H-1 \sim \text{Poisson}(\lambda = 1)$ \citep{geng2019probabilistic,hu2023bayesian}.

\subsection{Constrained Gibbs-type prior}
In the SRT data, the spots are constrained on a two-dimensional physical plane, resulting in neighboring spots being more likely to belong to the same spatial domain. Numerous statistical models adopt a MRF model on $\bm z$ to enhance the spatial coherence of neighboring spots \citep{li2022bass, yan2024bayesian, li2019bayesian}. The conditional distribution of cluster allocation $\bm z$ is 
\begin{equation}
    P(z_{i} = h | \bm{z}_{-i}) \propto \exp\left(g_h + d \sum_{i'=1}^n e_{ii'}{I}(z_{i'} = h)\right),
\end{equation}
where $g_h\in\mathbb{R}$ signifies the underlying mixing proportion of cluster $h$ and  $d\in\{0\}\cup\mathbb{R}^+$ controls the spatial dependency strength or spatial homogeneity among neighboring spots. Larger values of $d$ lead to high homogeneity in the clustering analysis, resulting in a smooth spatial domain pattern.  We can also write the joint MRF prior on $\bm{z}$ by
\begin{equation}
    P(\bm{z}) \propto \exp\left(\sum_{h=1}^{H}g_{h}\sum_{i=1}^{n}I(z_{i} = h) + d \sum_{i < i'}^n e_{ii'}{I}(z_{i} = z_{i'})\right).
\end{equation}

A solution to incorporate MRF constraint in the Gibbs-type prior relies on the product partition model properties, which is based on the idea of replacing Equation (\ref{Gibbs_prior}) with
\begin{equation}
    P(\bm{z}) \propto V_{n}(H) \prod_{h=1}^{H} \psi(\mathcal{G}^{(n)}_{h})(1 - \delta)_{n_{h}-1} .\label{eq:MRF_Gibbs_prior}
\end{equation}
Here we redefine the neighborhood relationship among spots \textit{via} undirected graph $\mathcal{G}^{(n)} = (\mathcal{V}^{(n)}, \mathcal{E}^{(n)})$ that consists of a set of nodes $\mathcal{V}^{(n)} = \{1, \ldots, n\}$ and a set of undirected edges $\mathcal{E}^{(n)} = \{e_{ii'}: i \in V^{(n)}, i' \in V^{(n)}\}$. In Equation \ref{eq:MRF_Gibbs_prior}, 
$\mathcal{G}^{(n)}_{h} = (\mathcal{V}^{(n)}_{h}, \mathcal E^{(n)}_{h})$ with $\mathcal V^{(n)}_{h} = \{i: z_{i} = h\}$ and $\mathcal E^{(n)}_{h} = \{e_{ii'}: i \in \mathcal V^{(n)}_{h}, i' \in \mathcal V^{(n)}_{h}\}$ is the subgraph of $\mathcal{G}$ and $\psi: \mathcal{G} \rightarrow \mathbb{R}^{+}$ is a mapping function from a graph to a non-negative value. We define it as the MRF-constrained Gibbs-type priors with $\psi(\mathcal{G}_{h}) = \exp(n_{h}g_{h} + d|\mathcal E^{(n)}_{h}|)$ where $|\mathcal E^{(n)}_{h}|$ represents the total number of edges in subgraph $\mathcal{G}^{(n)}$. The  MRF-constrained Gibbs-type priors lead to the following sequential urn scheme
\begin{equation}
    P(z_{i}=h\mid \bm{z}_{-i}) \propto 
\begin{cases}
\frac{\psi\left(\mathcal{G}_{h}^{(n)}\right)}{\psi\left(\mathcal{G}_{h}^{(n-1)}\right)}V_{n}(H)(n_{h, -i} - \delta) & \text{for } h = 1, \ldots H \\
V_{n}(H+1) & \text{for } h = H+1 \\
\end{cases}
\end{equation}
where $\bm{z}_{-i}$ denotes all entries in $\bm{z}$ excluding the $i$-th one and $n_{h,-i}$ is the number of spots in cluster $h$ ignoring the $i$-th one. When the newly added spots are connected with spots in existing cluster $h$, it has a higher probability of belonging to cluster $h$ if $d > 0$, which enforces the locally contiguous clusters. If the new spot is not connected with any spots or $d = 0$, the MRF-constrained Gibbs-type priors reduce to the vanilla Gibbs-type priors. According to \cite{li2010bayesian} and \cite{stingo2013integrative}, higher $d$ values may lead to phase a phase transition problem. To choose an appropriate value of $d$, we recommend using the \textit{integrated complete likelihood} (ICL) in Section \ref{subsec:model_selection}.

\section{Model Fitting}
\label{sec:infer}

\subsection{Model identifiability}
Non-identifiability of the latent structure creates problems in the interpretation of the results from
Markov chain Monte Carlo samples \citep{bhattacharya2011sparse, rovckova2016fast, schiavon2022generalized}. The parameter $\{\bm{W}, \Lambda, \{\bm \mu_{h}\}_{h\geq1}, \Sigma\}$ and latent variable $\bm{Y}$ in mdoel (\ref{eq:model}) is not identifiable. Indeed, for any nonsingular matrix $\bm{M} \in \mathbb{R}^{q \times q}$, the parameter $\{\bm{W}, \Lambda, \{\bm\mu_{h}\}_{h\geq1}, \Sigma\}$ and $\{\bm{W}\bm{M}^{-1}, \Lambda, \{\bm{M}\bm\mu_{h}\}_{h\geq 1} , \bm{M}\Sigma\bm{M}^{\top}\}$ lead to the same data likelihood. This issue generates, in practice, an infinite number of solutions for the parameter estimate. However, the cluster membership $\bm{z}$ is identifiable, which implies $\bm{z}$ is invariant under transformation matrix $\bm{M}$, as shown in Proposition \ref{prop:prop1} 

\begin{definition}
Let $\bm{Y}_{0} = (\bm{y}_{01}, \ldots, \bm{y}_{0n})$ be the true values of unobservable latent variables corresponding to $n$ observations. The following mixture model is assumed for each observation $i \in \{1, \cdots, n\}$ 
\begin{equation*}
    \bm{y}_{0i} \sim \sum_{h\geq 1} \pi_{h}\mathcal{K}(\bm{y}; \theta_{h}), \quad \theta_{h} \sim G_{0}, \quad \bm{\pi} \sim Q_{0},
\end{equation*}
then the posterior probability of observing the partition $\Psi$ induced by the clusters $z_{1}, \ldots, z_{n}$ conditionally on $\bm{Y}_{0}$ is
\begin{equation*}
    P(\Psi \mid \bm{Y}_{0}) =: \frac{P(\Psi) \times \int \prod_{h \geq 1} \prod_{i:z_{i}= h} \mathcal{K}(\bm{y}_{0i};\theta_{h}) dG_{0}(\theta_{h})} {\sum_{\Psi^{'}\in \mathcal{C}_{n}} P(\Psi^{'}) \times \int \prod_{h \geq 1} \prod_{i:z_{i}^{'}= h} \mathcal{K}(\bm{y}_{0i};\theta_{h}) dG_{0}(\theta_{h})}
\end{equation*}
where $\mathcal{C}_{n}$ is the space of all possible partitions of $n$ data points into clusters.
\end{definition}

\begin{proposition}
Under the following model specifications,
\begin{equation*}
    \mathcal{K}(\bm{y}; \theta_{h}) = N(\bm{y}; \bm\mu_{h}, \Sigma), \quad \bm\mu_{h} \mid \Sigma \sim N(\bm 0, \tau_{\,u}^{-1}\Sigma), \quad \Sigma \propto |\Sigma|^{-\frac{q+1}{2}}
\end{equation*}
with $q < n$, the probability $P(\Psi | \bm{Y}_{0}) = P(\Psi \mid \bm{MY}_{0})$ for any nonsingular matrix $\bm{M}$.  \label{prop:prop1}
\end{proposition}

\subsection{Posterior sampling}
In the model, $\bm{Y}$ and $\bm{z}$ are latent variables,  and $\Theta = \left\{\bm{W}, \Lambda, \{\bm\mu_{h}\}_{h=1}^{H}, \Sigma, H \right\}$ are model parameters. The posterior of latent variables and parameters is
\begin{equation*}
\begin{aligned}
    P(\bm{Y}, \bm{z}, \bm{W}, \Lambda, \{\bm \mu_{h}\}_{h=1}^{H}, \Sigma, H \mid \bm{X}) & \propto P(\bm{X \mid \bm{W}, \bm{Y}, \Lambda})P(\bm{Y} \mid \bm{z}, \{\bm\mu_{h}\}_{h=1}^{H}, \Sigma, H)\\
    & \times P(\bm{z} \mid H)P(\{\bm\mu_{h}\}_{h=1}^{H} \mid H, \Sigma)P(\Sigma)P(H).
\end{aligned}
\end{equation*}
For posterior computation, we adapt a collapsed Gibbs sampler \citep{neal2000markov} defined by the following steps.

\begin{itemize}
    \item Sample latent factors $\bm{y}_{i}$ for $i \in \{1, \ldots n\}$: The posterior of $\bm{y}_{i}$ is
    \begin{equation*}
        P(\bm{y}_{i}|z_{i} = h, -) \propto \exp\left[-\frac{1}{2}(\bm x_{i} - \bm{W}\bm y_{i})^{\top} \Lambda^{-1}(\bm x_{i} - \bm{W}\bm y_{i}) - \frac{1}{2}(\bm y_{i} - \bm\mu_{h})^{\top} \Sigma^{-1}(\bm y_{i} - \bm\mu_{h})\right].
    \end{equation*}
    Thus, draw $\bm{y}_{i}$ from normal distribution $N(\bm\mu^{*}, \Sigma^{*})$, where $\Sigma^{*} = (\bm{W}^{\top}\Lambda^{-1}\bm{W} + \Sigma^{-1})^{-1}$ and $\bm\mu^{*} = \Sigma^{*}(\bm{W}^{\top}\Lambda^{-1}\bm{x}_{i} + \Sigma^{-1}\bm\mu_{h})$. 

    \item Let $\bm{w}_{j\cdot}$ be the $j$-th row of $\bm{W}$, sample factor loadings $\bm{w}_{j\cdot}$ for $j \in \{1, \ldots, p\}$: The posterior of $\bm{w}_{j\cdot}$ is
    \begin{equation*}
        P(\bm{w}_{j\cdot} \mid -) \propto \exp\left[-\frac{1}{2}(\bm{x}_{j\cdot} - \bm{w}_{j\cdot} \bm{Y}) (\bm{x}_{j}^{\top} - \bm{w}_{j} \bm{Y})^{\top}/\sigma^{2}_{j}\right] \exp(-\frac{\tau_{w}||\bm{w}_{j\cdot}||^{2}}{2\sigma^{2}_{j}}).
    \end{equation*}
    Draw $\bm{w}_{j\cdot}$ from a normal distribution by $\bm{w}_{j\cdot} \sim N\left((\bm{Y}\bm{Y}^{\top} + \tau_{w} \bm{I}_{q})^{-1}\bm{Y}\bm{x}_{j\cdot}^{\top}, (\bm{Y}\bm{Y}^{\top} + \tau_{w} \bm{I}_{q})^{-1}\right)$ where $\bm{x}_{j\cdot}$ is the $j$-th row of $\bm{X}$. 

    \item Sample variance $\sigma^{2}_{j}$ for $j \in \{1, \ldots, p\}$: The posterior of $\sigma^{2}_{j}$ is
    \begin{equation*}
        P(\sigma^{2}_{j}\mid -) \propto \sigma_{j}^{-n-q} \exp\left[-\frac{1}{2\sigma^{2}_{j}}(\bm{x}_{j\cdot} - \bm{w}_{j\cdot} \bm{Y}) (\bm{x}_{j\cdot} - \bm{w}_{j\cdot} \bm{Y})^{\top}\right] \exp(-\frac{\tau_{w}||\bm{w}_{j\cdot}||^{2}}{2\sigma^{2}_{j}})\text{IG}(\sigma^{2}_{j}|a, b).
    \end{equation*}
    Draw $\sigma^{2}_{j}$ from inverse gamma distribution $\sigma^{2}_{j} \sim \text{IG}(a^{*}, b^{*})$, where $a^{*} = \frac{q+n}{2} + a$ and $b^{*} = \frac{1}{2}[(\bm{x}_{j\cdot} - \bm{w}_{j\cdot} \bm{Y}) (\bm{x}_{j\cdot} - \bm{w}_{j\cdot} \bm{Y})^{\top} + \tau_{w}||\bm{w}_{j\cdot}||^{2}] + b$, where $\bm x_{j\cdot}$ is the $j$-th row of $\bm{X}$. 

    \item Sample mean parameter $\bm \mu_{h}$ for $h \in \{1, \ldots H\}$: The posterior is
    \begin{equation*}
        P(\bm\mu_{h}\mid -) \propto \exp\left(-\frac{1}{2}\sum_{i:z_{i} = h}(\bm{y}_{i} - \bm\mu_{h})^{\top}\Sigma^{-1}(\bm{y}_{i} - \bm\mu_{h})\right) \exp(-\frac{1}{2}\tau_{\mu} \bm\mu_{h}^{\top}\Sigma^{-1}\bm\mu_{h}).
    \end{equation*}
    Draw $\bm\mu_{h}$ from normal distribution $\bm\mu_{h} \sim N\left(\frac{\sum_{i:z_{i} = h}\bm{y}_{i}}{n_{h} + \tau_{\mu}}, \frac{\Sigma}{n_{h} + \tau_{\mu}}\right)$

    \item Sample covariance $\Sigma$: 
    \begin{equation*}
        P(\Sigma \mid -) \propto \det(\Sigma)^{-\frac{n+H + q +1}{2}}\exp\left(-\frac{1}{2}\sum_{h=1}^{H}\sum_{i:z_{i} = h}(\bm{y}_{i} - \bm\mu_{h})^{\top}\Sigma^{-1}(\bm{y}_{i} - \bm\mu_{h}) -\frac{1}{2}\tau_{\mu} \sum_{h=1}^{H} \bm\mu_{h}^{\top}\Sigma^{-1}\bm\mu_{h}\right).
    \end{equation*}
    Sample $\Sigma$ from Inverse-Wishart distribution $\text{IW}(\bm{\Phi}, v)$ with $v = n+H$ and $\bm{\Phi} = \sum_{h=1}^{H}\left( \sum_{i, z_{i} = h}(\bm{y}_{i} - \bm\mu_{h})(\bm{y}_{i} - \bm\mu_{h})^{\top} + \tau_{\mu}\bm\mu_{h}\bm\mu_{h}^{\top}\right)$. 

    \item Sample cluster membership $\bm{z}$ sequentially: 
    \begin{equation*}
        \begin{aligned}
           & P(z_{i} = h \mid - )  \propto [n_{h, -i} + \beta]\exp\left(d \sum_{i' \neq i} e_{ii'}{I}(z_{i'} = h)\right) N(\boldsymbol{y}_{i}\mid \bm\mu_{h}, \Sigma) \quad \text{at existing cluster } h ,\\
           & P(z_{i} = H+1 \mid -) \propto \frac{V_{n}(H+1)}{V_{n}(H)} N\left(\bm{y}_{i} \mid 0, (1 + \frac{1}{\tau_{\mu}})\Sigma\right) \quad \text{for a new cluster }.
        \end{aligned}
    \end{equation*}
  
\end{itemize}

\subsection{Posterior inference on cluster membership}
For summarizing the posterior distribution of $\bm{z}$, we may use the \textit{maximum-a-posterior} (MAP) estimate
\begin{equation}
    \hat{\bm{z}}^{\text{MAP}}   =  \underset{1 \leq u \leq U }{\mathrm{argmax}}~ P\left(\bm{X}|\bm{W}^{(u)},\bm{Y}^{(u)},\Lambda^{(u)}\right)P(\bm{Y}^{(u)} |\bm{z}^{(u)}, \{\bm{\mu}_{h}^{(u)}\}_{h \geq 1}, \Sigma^{(u)})P\left(\bm{z}^{(u)}\right), 
\end{equation}
with $u=1, \ldots, U$ indicating the MCMC iterations, after burn-in. In this paper, we recommend obtaining a summary of $\bm{z}$ based on the pairwise probability matrix (PPM) \citep{dahl2006model}. The PPM, an $n\times n$ symmetric matrix, calculates the posterior pairwise probabilities of co-clustering; that is, the probability that spot $i$ and $i'$ are assigned to the same cluster: $\text{PPM}_{i,i'}\approx\sum_{u=1}^U{I}(z_i^{(u)}=z_{i'}^{(u)})/U$. A point estimate $\hat{\bm z}^{\text{PPM}}$ can then be derived by minimizing the sum of squared deviations between its association matrix and the PPM:
\begin{equation}
    \hat{\bm z}^{\text{PPM}}  =  \underset{\bm z \in \{\bm z^{(1)}, \ldots, \bm z^{(U)}\}}{\mathrm{argmax}}~\sum_{i<i'} \big({I}({z}_i={z}_{i'})-\text{PPM}_{ii'}\big)^2.
\end{equation}
The PPM estimate has the advantage of utilizing information from all clusterings through the PPM. It is also intuitively appealing because it selects the `average' clustering rather than forming a clustering \textit{via} an external, \textit{ad hoc} clustering algorithm. This approach ensures a more comprehensive and representative summary of the clustering outcomes. 

\subsection{Model selection}
\label{subsec:model_selection}
In the MRF-constrained Gibbs-type model, choosing the hyperparameter $d$ in the MRF is important, which controls the magnitude of spatial smoothness. Higher $d$ leads to higher spatial smoothness, resulting in a smaller number of clusters and decreasing the model complexity. We recommended using the \textit{integrated complete likelihood} (ICL) \citep{biernacki2000assessing} to find an appropriate $d$. The ICL is calculated as
\begin{equation}
    \text{ICL}(d) = -2 \log\left(L(\bm{X}, \hat{\bm{Y}}, \hat{\bm{z}} \mid \hat{\bm{W}}, \hat{\Lambda}, \{\hat{\bm\mu}_{h}\}_{h \geq 1}, \hat{\Sigma}, d )\right) + \log(n)(p \times q + \hat{H} \times q + q(q+1)/2 + p), 
\end{equation}
where $\hat{H}$ is the estimated number of estimated clusters, $\hat{\bm{Y}}$ and $\hat{\bm{z}}$ are estimated latent low-dimensional embedddings and cluster allocations. In the formula, the complete likelihood is given by
\begin{equation*}
\begin{aligned}
     & L(\bm{X}, \hat{\bm{Y}}, \hat{\bm{z}}\mid \hat{\bm{W}}, \{\hat{\bm\mu}_{h}\}_{h \geq 1}, \hat{\Sigma}, d ))  = P(\bm{X}|\hat{\boldsymbol{Y}}, \hat{\bm{W}}, \hat{\Lambda}) P(\hat{\bm{Y}}|\hat{\bm{z}}, \{\hat{\bm\mu}_{h}\}_{h \geq 1}, \hat{\Sigma})P(\hat{\bm{z}}) \\
     & \propto \prod_{i=1}^{n} \prod_{h=1}^{\hat{H}}N(\bm{x}_{i} \mid \hat{\bm{W}} \hat{\bm{y}}_{i}, \hat{\Lambda})\left[N(\hat{\bm{y}_{i}} \mid \hat{\bm\mu}_{h}, \hat{\Sigma})\right]^{1(\hat{z}_{i} = h)} V_{n}(\hat{H}) \prod_{h=1}^{\hat{H}}\psi(\mathcal{G}^{(n)}_{h})(1 - \delta)_{n_{h}-1}.
\end{aligned}
\end{equation*}
where the estimators are the values in the last iteration of the MCMC samples.

\section{Simulation Study}
\label{sec:sim}
To assess the performance of BNPMFA and quantify the advantages over state-of-the-art alternatives, we generated simulated SRT datasets based on two artificial spatial patterns and one real spatial domain pattern from the human DLPFC 10x Visium data. The first two patterns, displayed in Figure S1 of Supplementary Material as Patterns I and II, are on an $n=40\times40$ square lattice. Their $\bm{z}$'s were generated from Potts models with the true number of clusters $H_{0}=3$ and $5$ states, respectively, using the \verb|sampler.mrf| function in the \verb|R| package \verb|GiRaF|. The real spatial pattern, illustrated as Pattern III in Figure S1, contains $n=4,266$ spots across $H_{0}=7$ spatial domains. The latent expression $\bm{y}_{i}$ is generated from Gaussian distribution condition on $z_{i} = h$ via $\bm y_{i} \mid z_{i} = h \sim N(\mu_{h}, \Sigma)$ with latent dimension $q = 10$ for $i \in \{1, \ldots, n\}$ and $ h \in \{1, \ldots, H_{0}\}$. The factor loadings $\bm{W} = [w_{jl}]_{p \times q}$ are sampled from standard normal distribution by $w_{jl} \sim N(0, 1)$ with $p = 2000$ features. Finally, $\bm{x}_{i} = \bm{W} \bm{y}_{i} + \bm{\epsilon}_{i}$, where $\bm{\epsilon}_{i} \sim N(0, \text{diag}(\sigma^{2}_{1}, \ldots, \sigma^{2}_{p}))$ with $\sigma^{2}_{j} \sim \text{IG}(2, 1)$. For the parameter, $\bm\mu_{h}$ and $\Sigma$, we consider two cases, strong signal, and weak signal. The detailed parameter settings are in Table S1 of the Supplementary Material.

To assess the performance of different Gibbs-type priors, we implemented three representative priors: PY, DP, and MFM, along with their MRF-constrained versions, in various simulation scenarios. In the hyperparameter settings, we fixed $\beta = 1$ for all Gibbs-type priors and set $\delta = -0.1$ for both PY and MRF-constrained PY. Regarding the MRF, we set $g_{1}= \cdots g_{H} = 1$ to ensure noninformative priors on cluster abundance. The spatial smooth parameter $d$ was determined using the ICL criterion, with a detailed estimation provided for each scenario in Table S2 of the Supplementary Material. For the hyperparameters in the priors of $\bm{W}, \{\bm\mu_{h}\}_{h \geq 1}, \Lambda$, we set $\tau_{u} = \tau_{w} = 1$ and $a = b =1$. To evaluate the clustering performance through $\boldsymbol{z}$ across various methods, we used the adjusted Rand index (ARI). The ARI, which ranges from 0 to 1, measures the similarity between two different partitions. Higher ARI values indicate more accurate clustering outcomes, with a value of one indicating a perfect match. 

Among the Gibbs-type prior models without an MRF constraint, the MFM model consistently exhibited the best clustering performance across the majority of simulation scenarios. Specifically, the MRF-constrained PY, DP, and MFM achieved similar clustering performance and identified the true number of clusters under strong signal, but under weak signal case, the MRF-constrained MFM model achieved mean cluster numbers closest to the ground truth (Table \ref{tab:sim1_ari} and Table \ref{tab:sim1_H}). The MRF constraint exhibits a higher contribution to the clustering process, leading to similar performance for different Gibbs-type priors. The simulation results correspond to the MFM model property that the MFM model can achieve a consistent number of cluster estimations when the data are generated from a finite mixture distribution.

\begin{figure}
\begin{center}
\includegraphics[width = 1\textwidth]{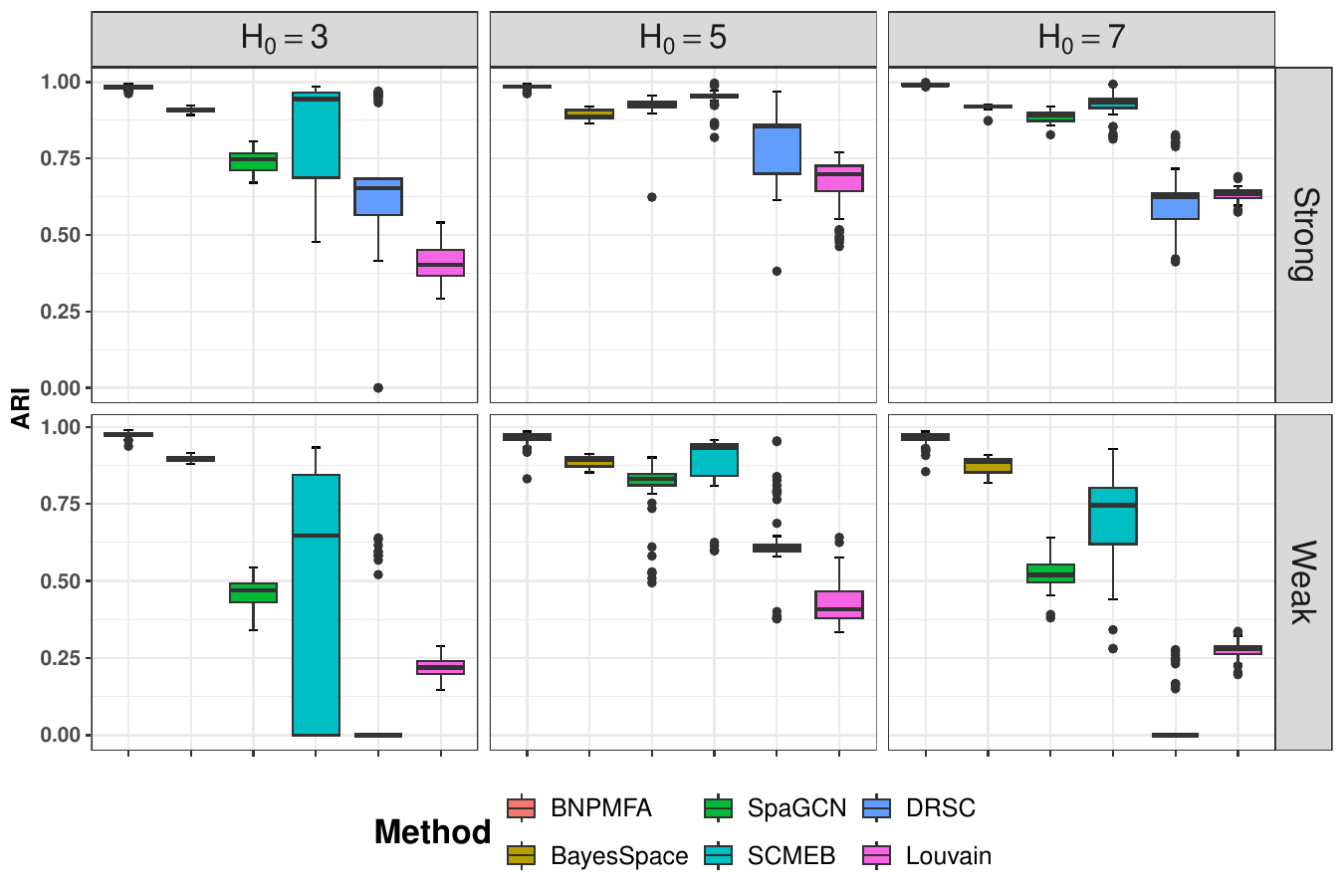}
\end{center}
\caption{Simulated data analysis: Boxplots of ARIs achieved by BNPMFA and competing methods across various scenarios in terms of spatial domain patterns and signal settings settings.  
\label{fig:sim_ari}}
\end{figure}

To further clarify the magnitude of the improvements of the proposed method, Figure \ref{fig:sim_ari} illustrates the performance of the MRF-constrained MFM model (BNPMFA) and the state-of-the-art alternatives, including SCMEB \citep{yang2022sc}, DRSC \citep{liu2022joint}, BayesSpace \citep{zhao2021spatial}, SpaGCN \citep{hu2021spagcn}, and Louvain. The detailed implementation settings of competing methods are provided in the Supplementary Material. BNPMFA achieved the best clustering performance in all simulation scenarios. For the number of cluster estimation, only SCMEB and DRSC methods utilize the MBIC criterion as a post-selection strategy to determine the number of clusters among the competing methods. As illustrated in Figure S4, MBIC utilized by SCMEB and DRSC displays a tendency to systematically under-estimate the true number of clusters and these methods exhibit reduced accuracy in learning true partition. BNPMFA achieved the best performance in both the clustering task and the estimation on the number of clusters.

To evaluate and compare the robustness of BNPMFA and competing methods, we simulated spatially resolved transcriptomic data sets by scDesign3 \citep{song2024scdesign3} based on the BZ5 STARmap data utilized in Section \ref{subsec:STARmap}. Details are in supplementary Section S2.3, which indicates that BNPMFA is more robust than competing methods to the scDesign3-induced SRT data.

\section{Case Study}
\label{sec:app}

\subsection{Application to the Human DLPFC 10X Visium Data}
\label{subsec:DLPFC}

\begin{figure}
\begin{center}
\includegraphics[width = 1\textwidth]{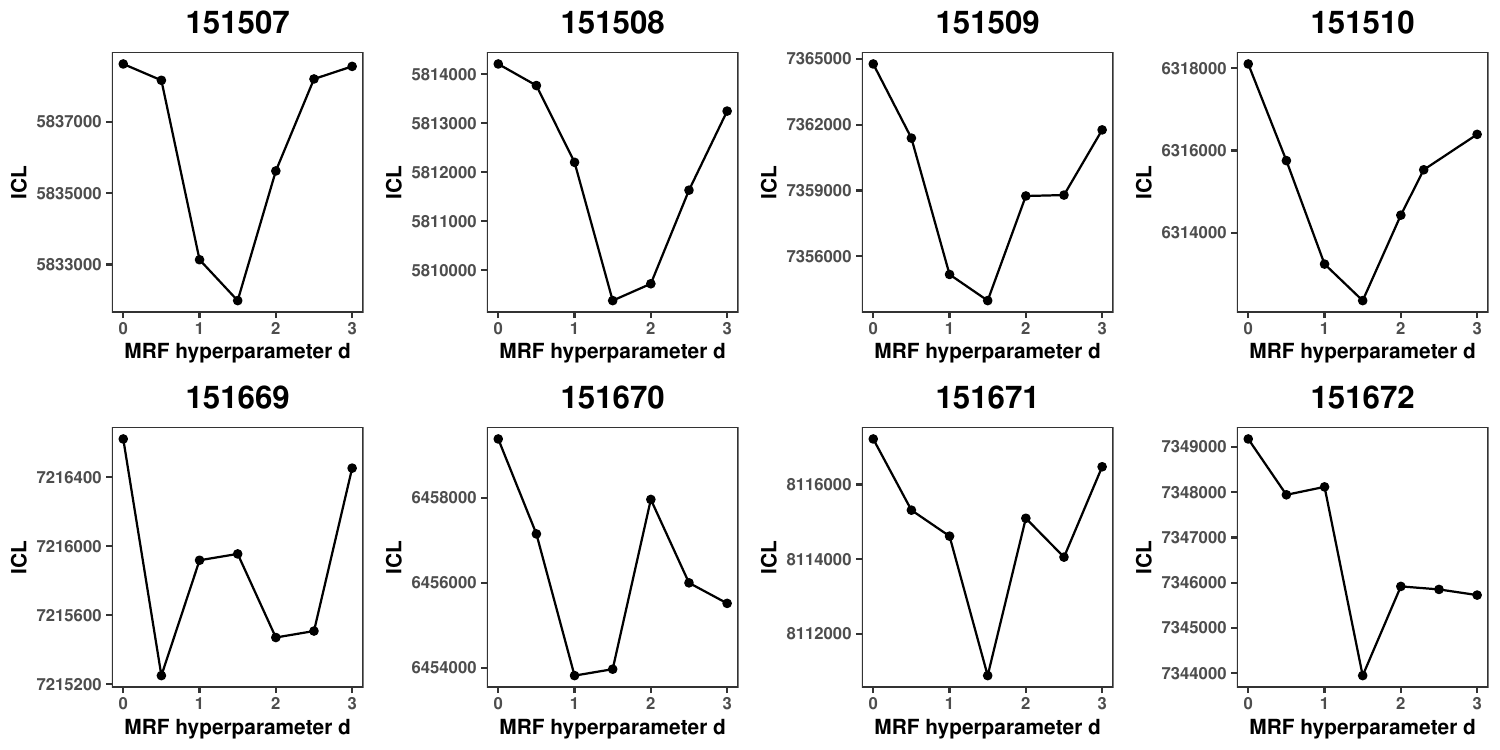}
\end{center}
\caption{DLPFC data analysis: MRF hyperparameter $d$ for BNPMFA in $8$ DLPFC samples.  \label{fig:dlpfc_model_selection}}
\end{figure}

\cite{maynard2021transcriptome} utilized the 10x Visium assay to generate spatial molecular profiles for human dorsolateral prefrontal cortex (DLPFC) samples, which contains a total of $33,538$ genes on approximate $4000$ spots with triangle lattice structure. They manually annotated $6$ or $4$ cortical layers and white matter based on cytoarchitecture, resulting in $H_{0}=7$ or $H_{0} = 5$ annotated spatial domains, respectively. The data is accessible through the \verb|R| package \verb|spatialLIBD| \citep{pardo2022spatiallibd} and is extensively benchmarked in SRT methodological research. Following the preprocessing procedure widely adopted in SRT data analysis \citep{zhao2021spatial,yang2022sc, yan2024bayesian}, we use the log-normalization method in R package \verb|scran| and keep the top $2000$ highly variable genes (HVGs) as the input of the BNPMFA model with $q = 15$ latent dimensions. For the MRF hyperparameter, we selected optimal $d$ informed by the ICL plot shown in Figure \ref{fig:dlpfc_model_selection} under $8$ tissue samples. We initiated four independent MCMC chains and calculated the pairwise ARIs between the estimation of cluster allocation $\bm z$ for different chains. The ARIs range from $0.925$ to $0.982$, indicating a good mixing of our MCMC algorithm. 

\begin{figure}
\begin{center}
\includegraphics[width = 1.2\textwidth,angle =90]{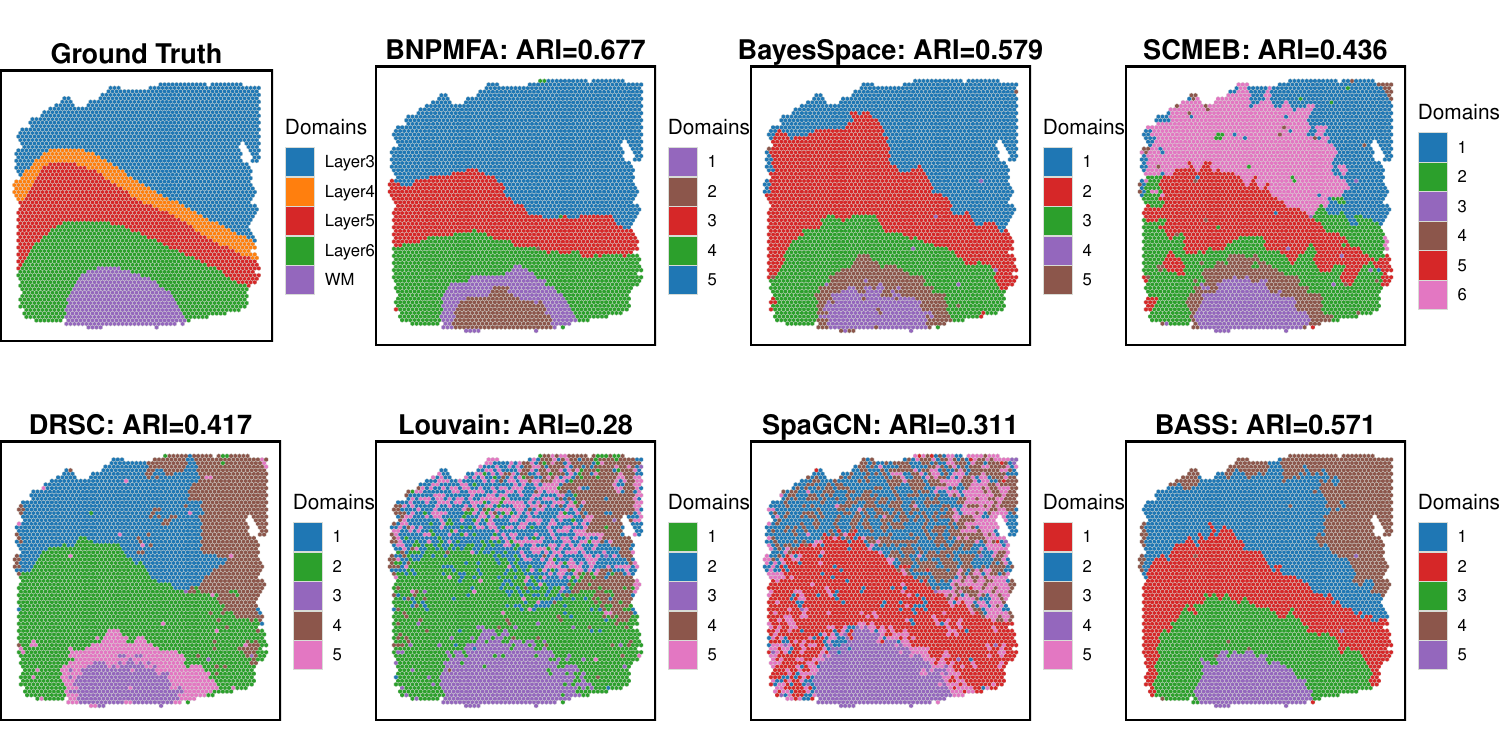}
\end{center}
\caption{DLPFC data analysis: Spatial domains annotated by pathologists and
identified by BNPMFA and competing methods in DLPFC sample 151672.}  \label{fig:151672_cluster}
\end{figure}

Here we evaluate BNPMFA’s ability to identify distinct layer-specific expression profiles and compare its performance to other state-of-the-art spatial clustering alternatives under 8 tissue samples. BNPMFA substantially outperforms the competing spatial clustering methods and yields the best clustering performance in 6 tissue samples out of a total of 8 samples (Table \ref{tab:dlpfc_ari} and Figure S5-S11). As an example, in sample 151672 (Figure \ref{fig:151672_cluster}), methods (BNPMFA, BayesSpace, SCMEB, DRSC, BASS) incorporating MRF identified continuous spatial domains, indicating the contribution of MRF in the spatial clustering frameworks. For the number of clusters estimation, BNPMFA discovers the true number of clusters in 6 out 8 samples, as shown in Table S3 of Supplementary Material. 

\subsection{Application to the Mouse STARmap Data}
\label{subsec:STARmap}

\begin{figure}
\begin{center}
\includegraphics[width = 1\textwidth]{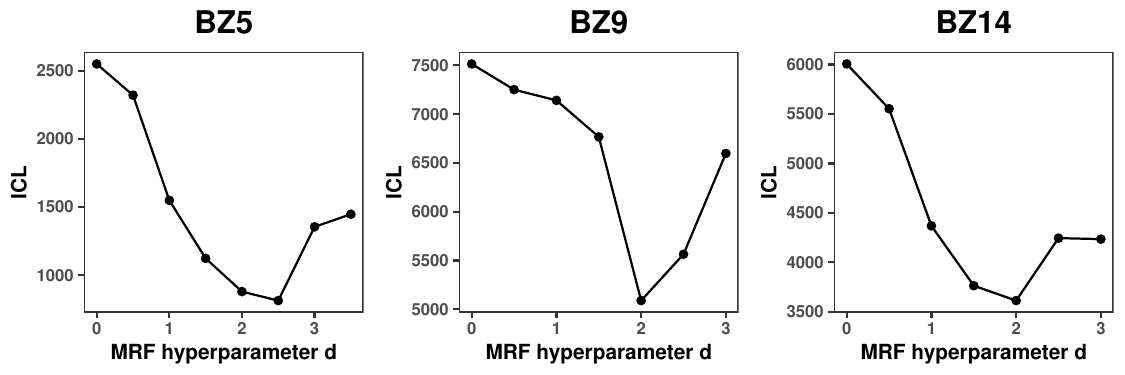}
\end{center}
\caption{STARmap data analysis: MRF hyperparameter $d$ for BNPMFA in $3$ STARmap samples.  \label{fig:starmap_model_selection}}
\end{figure}

The second dataset is from the STARmap technology, consisting of three tissue sections obtained from the medial prefrontal cortex (mPFC) of the mouse brain from different mice \citep{wang2018three}. The data set contains three tissue slices including BZ5 (1049 cells), BZ9 (1053 cells), and BZ14 (1088 cells), with expression measurements collected on a common set of 166 genes. Among the 166 genes, 112 of them are putative cell-type markers and 48 of them are activity-regulated genes. The cells on all tissue sections were carefully annotated to four distinct layer structures that included L1, L2/3, L5, and L6 \citep{li2022bass}. For the preprocessing procedure, the log-normalized gene expression data as the input of BNPMFA and competing methods. For the MRF hyperparameter, we selected optimal $d$ informed by the ICL plot shown in Figure \ref{fig:starmap_model_selection} under $3$ tissue samples. We initiated four independent MCMC chains and calculated the pairwise ARIs between the estimation of $\bm z$ for different chains. The ARIs range from 0.894 to 0.952, indicating a good mixing of our MCMC algorithm.

\begin{figure}
\begin{center}
\includegraphics[width = 1\textwidth]{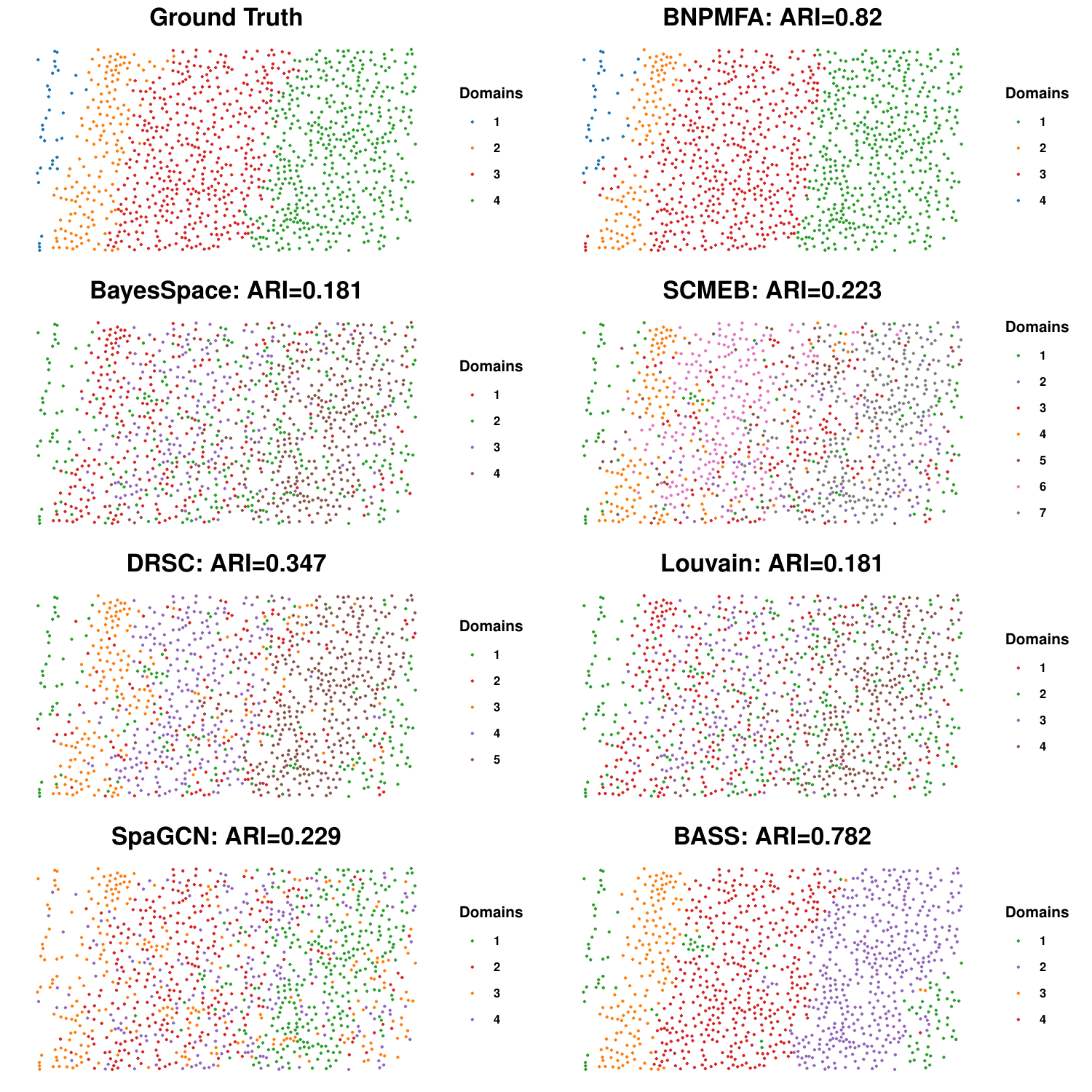}
\end{center}
\caption[Spatial domains of STARmap BZ14 sample]{STARmap data analysis: Spatial domains annotated by pathologists and
identified by BNPMFA and competing methods in STARmap sample BZ14.}  \label{fig:BZ14_cluster}
\end{figure}

Table \ref{tab:dlpfc_ari} provides the clustering performance of BNPMFA and alternatives under 3 tissue slices where BNPMFA achieved the best performance in 2 tissue slices. As shown in the results of BZ14 (Figure \ref{fig:BZ14_cluster}), BNPMFA realized a clearer boundary between domains compared with alternative methods, leading to smoother and biologically meaningful domains. The second output of our model is the estimation of the number of domains (clusters). BNPMFA estimated mean cluster number 4 in three tissues, which is equal to the annotated number of clusters. The competing methods, SCMEB and DRSC, tend to overestimate the number of clusters. In summary, BNPMFA achieved the best clustering performance without a pre-specified number of clusters and provided a good estimation for the number of clusters. 

\section{Conclusion}
\label{sec:con}

In this paper, we introduce a novel class of Bayesian NonParametric Mixture of Factor Analysis (BNPMFA) models designed to address the high-dimensional challenges inherent in spatial omics data analysis. By leveraging factor analysis, our approach projects high-dimensional observational data into a lower-dimensional space, facilitating more manageable and interpretable analysis. These models integrate a MRF-constrained Gibbs-type prior on the partition process, effectively incorporating spatial dependencies within SRT data. This novel prior enables the simultaneous inference of cluster membership and the determination of the optimal number of clusters, a significant advancement in the SRT data clustering analysis

Within this framework, we have rigorously established the mathematical identifiability of cluster membership, ensuring the robustness of our approach. To facilitate inference, we developed an efficient MCMC sampling algorithm that combines Gibbs sampling with a collapsed Gibbs sampler tailored for Bayesian nonparametric models. The efficacy of our proposed BNPMFA model is validated through extensive simulations and applied to two benchmark SRT datasets. Our results demonstrate that BNPMFA outperforms state-of-the-art methods in clustering accuracy, as measured by the ARI, and in estimating the number of clusters and cluster configurations. Moreover, our proposed method offers fresh insights and a deeper understanding of cellular regions within tissue samples.

From a computational perspective, our method exhibits linear computational time complexity with respect to both the feature dimension $p$ and the sample size $n$, as detailed in Supplementary Section S2.2. Specifically, the computational complexity is approximately $\mathcal{O}(pqU + nqHU)$, highlighting the scalability of our algorithm for large-scale datasets.

Nevertheless, it is important to acknowledge the limitations of assuming a Gaussian distribution in the factor analysis model, which may not fully capture the characteristics of sequence count data, particularly its often non-Gaussian nature. In future work, we aim to extend the BNPMFA framework to accommodate alternative distributions, such as the zero-inflated negative binomial distribution, which is more appropriate for modeling sequence count data in omics studies. Additionally, while our current approach uses criterion-based tuning for the MRF hyperparameter 
$d$, future research could explore the integration of tuning-free algorithms, such as placing a prior on $d$, to enhance model efficiency and reduce reliance on manual tuning.

\bibliographystyle{agsm}
{\bibliography{ref}}

@article{marx2021method,
  title={Method of the Year: spatially resolved transcriptomics},
  author={Marx, Vivien},
  journal={Nature Methods},
  volume={18},
  number={1},
  pages={9--14},
  year={2021},
  publisher={Nature Publishing Group US New York}
}

@inproceedings{zhao2023spatial,
  title={Spatial clustering regression of count value data via bayesian mixture of finite mixtures},
  author={Zhao, Peng and Yang, Hou-Cheng and Dey, Dipak K and Hu, Guanyu},
  booktitle={Proceedings of the 29th ACM SIGKDD Conference on Knowledge Discovery and Data Mining},
  pages={3504--3512},
  year={2023}
}

@article{meng2024bayesian,
  title={Bayesian Spatially Clustered Compositional Regression: Linking intersectoral GDP contributions to Gini Coefficients},
  author={Meng, Jingcheng and Ren, Yimeng and Zhu, Xuening and Hu, Guanyu},
  journal={arXiv preprint arXiv:2405.07408},
  year={2024}
}

@article{lubeck2014single,
  title={Single-cell in situ {RNA} profiling by sequential hybridization},
  author={Lubeck, Eric and Coskun, Ahmet F and Zhiyentayev, Timur and Ahmad, Mubhij and Cai, Long},
  journal={Nature Methods},
  volume={11},
  number={4},
  pages={360--361},
  year={2014},
  publisher={Nature Publishing Group US New York}
}

@article{chen2015spatially,
  title={Spatially resolved, highly multiplexed {RNA} profiling in single cells},
  author={Chen, Kok Hao and Boettiger, Alistair N and Moffitt, Jeffrey R and Wang, Siyuan and Zhuang, Xiaowei},
  journal={Science},
  volume={348},
  number={6233},
  pages={aaa6090},
  year={2015},
  publisher={American Association for the Advancement of Science}
}

@article{wang2018three,
  title={Three-dimensional intact-tissue sequencing of single-cell transcriptional states},
  author={Wang, Xiao and Allen, William E and Wright, Matthew A and Sylwestrak, Emily L and Samusik, Nikolay and Vesuna, Sam and Evans, Kathryn and Liu, Cindy and Ramakrishnan, Charu and Liu, Jia and others},
  journal={Science},
  volume={361},
  number={6400},
  pages={eaat5691},
  year={2018},
  publisher={American Association for the Advancement of Science}
}

@article{rodriques2019slide,
  title={Slide-seq: {A} scalable technology for measuring genome-wide expression at high spatial resolution},
  author={Rodriques, Samuel G and Stickels, Robert R and Goeva, Aleksandrina and Martin, Carly A and Murray, Evan and Vanderburg, Charles R and Welch, Joshua and Chen, Linlin M and Chen, Fei and Macosko, Evan Z},
  journal={Science},
  volume={363},
  number={6434},
  pages={1463--1467},
  year={2019},
  publisher={American Association for the Advancement of Science}
}

@article{staahl2016visualization,
  title={Visualization and analysis of gene expression in tissue sections by spatial transcriptomics},
  author={St{\aa}hl, Patrik L and Salm{\'e}n, Fredrik and Vickovic, Sanja and Lundmark, Anna and Navarro, Jos{\'e} Fern{\'a}ndez and Magnusson, Jens and Giacomello, Stefania and Asp, Michaela and Westholm, Jakub O and Huss, Mikael and others},
  journal={Science},
  volume={353},
  number={6294},
  pages={78--82},
  year={2016},
  publisher={American Association for the Advancement of Science}
}

@article{vickovic2019high,
  title={High-definition spatial transcriptomics for in situ tissue profiling},
  author={Vickovic, Sanja and Eraslan, G{\"o}kcen and Salm{\'e}n, Fredrik and Klughammer, Johanna and Stenbeck, Linnea and Schapiro, Denis and {\"A}ij{\"o}, Tarmo and Bonneau, Richard and Bergenstr{\aa}hle, Ludvig and Navarro, Jos{\'e} Fernand{\'e}z and others},
  journal={Nature Methods},
  volume={16},
  number={10},
  pages={987--990},
  year={2019},
  publisher={Nature Publishing Group US New York}
}

@article{zhang2021spatial,
  title={Spatial molecular profiling: platforms, applications and analysis tools},
  author={Zhang, Minzhe and Sheffield, Thomas and Zhan, Xiaowei and Li, Qiwei and Yang, Donghan M and Wang, Yunguan and Wang, Shidan and Xie, Yang and Wang, Tao and Xiao, Guanghua},
  journal={Briefings in Bioinformatics},
  volume={22},
  number={3},
  pages={bbaa145},
  year={2021},
  publisher={Oxford University Press}
}

@article{thrane2018spatially,
  title={Spatially resolved transcriptomics enables dissection of genetic heterogeneity in stage {III} cutaneous malignant melanoma},
  author={Thrane, Kim and Eriksson, Hanna and Maaskola, Jonas and Hansson, Johan and Lundeberg, Joakim},
  journal={Cancer Research},
  volume={78},
  number={20},
  pages={5970--5979},
  year={2018},
  publisher={AACR}
}

@article{moses2022museum,
  title={Museum of spatial transcriptomics},
  author={Moses, Lambda and Pachter, Lior},
  journal={Nature Methods},
  volume={19},
  number={5},
  pages={534--546},
  year={2022},
  publisher={Nature Publishing Group US New York}
}

@article{zhao2021spatial,
  title={Spatial transcriptomics at subspot resolution with {BayesSpace}},
  author={Zhao, Edward and Stone, Matthew R and Ren, Xing and Guenthoer, Jamie and Smythe, Kimberly S and Pulliam, Thomas and Williams, Stephen R and Uytingco, Cedric R and Taylor, Sarah EB and Nghiem, Paul and others},
  journal={Nature Biotechnology},
  volume={39},
  number={11},
  pages={1375--1384},
  year={2021},
  publisher={Nature Publishing Group US New York}
}

@article{yang2022sc,
  title={{SC-MEB}: {Spatial} clustering with hidden {Markov} random field using empirical {Bayes}},
  author={Yang, Yi and Shi, Xingjie and Liu, Wei and Zhou, Qiuzhong and Chan Lau, Mai and Chun Tatt Lim, Jeffrey and Sun, Lei and Ng, Cedric Chuan Young and Yeong, Joe and Liu, Jin},
  journal={Briefings in Bioinformatics},
  volume={23},
  number={1},
  pages={bbab466},
  year={2022},
  publisher={Oxford University Press}
}

@article{liu2022joint,
  title={Joint dimension reduction and clustering analysis of single-cell {RNA-seq} and spatial transcriptomics data},
  author={Liu, Wei and Liao, Xu and Yang, Yi and Lin, Huazhen and Yeong, Joe and Zhou, Xiang and Shi, Xingjie and Liu, Jin},
  journal={Nucleic Acids Research},
  volume={50},
  number={12},
  pages={e72--e72},
  year={2022},
  publisher={Oxford University Press}
}

@article{hu2021spagcn,
  title={{SpaGCN}: {Integrating} gene expression, spatial location and histology to identify spatial domains and spatially variable genes by graph convolutional network},
  author={Hu, Jian and Li, Xiangjie and Coleman, Kyle and Schroeder, Amelia and Ma, Nan and Irwin, David J and Lee, Edward B and Shinohara, Russell T and Li, Mingyao},
  journal={Nature Methods},
  volume={18},
  number={11},
  pages={1342--1351},
  year={2021},
  publisher={Nature Publishing Group US New York}
}

@article{li2023interpretable,
	title={An Interpretable {B}ayesian Clustering Approach with Feature Selection for Analyzing Spatially Resolved Transcriptomics Data},
	author={Li, Huimin and Zhu, Bencong and Jiang, Xi and Guo, Lei and Xie, Yang and Xu, Lin and Li, Qiwei},
	journal={Biometrics},
volume={80},
  number={3},
	pages={ujae066},
	year={2024},
	publisher={Oxford University Press}
}

@article{zhu2023bayesian,
  title={Bayesian Nonparametric Clustering with Feature Selection for Spatially Resolved Transcriptomics Data},
  author={Zhu, Bencong and Hu, Guanyu and Xie, Yang and Xu, Lin and Fan, Xiaodan and Li, Qiwei},
  journal={arXiv preprint arXiv:2312.08324},
  year={2023}
}

@article{li2022bass,
  title={{BASS}: multi-scale and multi-sample analysis enables accurate cell type clustering and spatial domain detection in spatial transcriptomic studies},
  author={Li, Zheng and Zhou, Xiang},
  journal={Genome Biology},
  volume={23},
  number={1},
  pages={168},
  year={2022},
  publisher={Springer}
}

@article{chandra2023escaping,
  title={Escaping the curse of dimensionality in {Bayesian} model-based clustering},
  author={Chandra, Noirrit Kiran and Canale, Antonio and Dunson, David B},
  journal={Journal of Machine Learning Research},
  volume={24},
  number={144},
  pages={1--42},
  year={2023}
}

@article{legramanti2022extended,
  title={Extended stochastic block models with application to criminal networks},
  author={Legramanti, Sirio and Rigon, Tommaso and Durante, Daniele and Dunson, David B},
  journal={The Annals of Applied Statistics},
  volume={16},
  number={4},
  pages={2369--2395},
  year={2022},
  publisher={NIH Public Access}
}

@article{miller2018mixture,
  title={Mixture models with a prior on the number of components},
  author={Miller, Jeffrey W and Harrison, Matthew T},
  journal={Journal of the American Statistical Association},
  volume={113},
  number={521},
  pages={340--356},
  year={2018},
  publisher={Taylor \& Francis}
}

@article{li2017bayesian,
  title={A {Bayesian} mixture model for clustering and selection of feature occurrence rates under mean constraints},
  author={Li, Qiwei and Guindani, Michele and Reich, Brian J and Bondell, Howard D and Vannucci, Marina},
  journal={Statistical Analysis and Data Mining: The ASA Data Science Journal},
  volume={10},
  number={6},
  pages={393--409},
  year={2017},
  publisher={Wiley Online Library}
}

@article{zhu2018identification,
  title={Identification of spatially associated subpopulations by combining {scRNAseq} and sequential fluorescence in situ hybridization data},
  author={Zhu, Qian and Shah, Sheel and Dries, Ruben and Cai, Long and Yuan, Guo-Cheng},
  journal={Nature Biotechnology},
  volume={36},
  number={12},
  pages={1183--1190},
  year={2018},
  publisher={Nature Publishing Group}
}

@article{markos2019beyond,
  title={Beyond tandem analysis: {Joint} dimension reduction and clustering in {R}},
  author={Markos, Angelos and D'Enza, Alfonso Iodice and van de Velden, Michel},
  journal={Journal of Statistical Software},
  volume={91},
  pages={1--24},
  year={2019}
}

@article{ferguson1973bayesian,
  title={A {Bayesian} analysis of some nonparametric problems},
  author={Ferguson, Thomas S},
  journal={The Annals of Statistics},
  volume = {1},
  pages={209--230},
  year={1973},
  publisher={JSTOR}
}

@article{fruhwirth2021generalized,
  title={Generalized mixtures of finite mixtures and telescoping sampling},
  author={Fr{\"u}hwirth-Schnatter, Sylvia and Malsiner-Walli, Gertraud and Gr{\"u}n, Bettina},
  journal={Bayesian Analysis},
  volume={16},
  number={4},
  pages={1279--1307},
  year={2021},
  publisher={International Society for Bayesian Analysis}
}

@article{de2013gibbs,
  title={Are {Gibbs-type} priors the most natural generalization of the {Dirichlet} process?},
  author={De Blasi, Pierpaolo and Favaro, Stefano and Lijoi, Antonio and Mena, Rams{\'e}s H and Pr{\"u}nster, Igor and Ruggiero, Matteo},
  journal={IEEE Transactions on Pattern Analysis and Machine Intelligence},
  volume={37},
  number={2},
  pages={212--229},
  year={2013},
  publisher={IEEE}
}

@article{camerlenghi2024contaminated,
  title={Contaminated {Gibbs-type} priors},
  author={Camerlenghi, Federico and Corradin, Riccardo and Ongaro, Andrea},
  journal={Bayesian Analysis},
  volume={19},
  number={2},
  pages={347--376},
  year={2024},
  publisher={International Society for Bayesian Analysis}
}

@article{hu2023bayesian,
  title={Bayesian spatial homogeneity pursuit of functional data: an application to the {U.S.} income distribution},
  author={Hu, Guanyu and Geng, Junxian and Xue, Yishu and Sang, Huiyan},
  journal={Bayesian Analysis},
  volume={18},
  number={2},
  pages={579--605},
  year={2023},
  publisher={International Society for Bayesian Analysis}
}

@article{geng2019probabilistic,
  title={Probabilistic community detection with unknown number of communities},
  author={Geng, Junxian and Bhattacharya, Anirban and Pati, Debdeep},
  journal={Journal of the American Statistical Association},
  volume={114},
  number={526},
  pages={893--905},
  year={2019},
  publisher={Taylor \& Francis}
}

@article{yan2024bayesian,
  title={Bayesian Integrative Region Segmentation in Spatially Resolved Transcriptomic Studies},
  author={Yan, Yinqiao and Luo, Xiangyu},
  journal={Journal of the American Statistical Association},
  number={just-accepted},
  pages={1--21},
  year={2024},
  publisher={Taylor \& Francis}
}

@techreport{ghahramani1996algorithm,
  title={The {EM} algorithm for mixtures of factor analyzers},
  author={Ghahramani, Zoubin and Hinton, Geoffrey E and others},
  year={1996},
  institution={Technical Report CRG-TR-96-1, University of Toronto}
}

@article{baek2009mixtures,
  title={Mixtures of factor analyzers with common factor loadings: Applications to the clustering and visualization of high-dimensional data},
  author={Baek, Jangsun and McLachlan, Geoffrey J and Flack, Lloyd K},
  journal={IEEE Transactions on Pattern Analysis and Machine Intelligence},
  volume={32},
  number={7},
  pages={1298--1309},
  year={2009},
  publisher={IEEE}
}

@article{lijoi2007bayesian,
  title={Bayesian nonparametric estimation of the probability of discovering new species},
  author={Lijoi, Antonio and Mena, Rams{\'e}s H and Pr{\"u}nster, Igor},
  journal={Biometrika},
  volume={94},
  number={4},
  pages={769--786},
  year={2007},
  publisher={Oxford University Press}
}

@article{miller2013simple,
  title={A simple example of {Dirichlet} process mixture inconsistency for the number of components},
  author={Miller, Jeffrey W and Harrison, Matthew T},
  journal={Advances in Neural Information Processing Systems},
  volume={26},
  pages = {199-206},
  year={2013}
}

@article{li2019bayesian,
  title={A {Bayesian} mark interaction model for analysis of tumor pathology images},
  author={Li, Qiwei and Wang, Xinlei and Liang, Faming and Xiao, Guanghua},
  journal={The Annals of Applied Statistics},
  volume={13},
  number={3},
  pages={1708--1732},
  year={2019},
  publisher={NIH Public Access}
}

@article{bhattacharya2011sparse,
  title={Sparse {Bayesian} infinite factor models},
  author={Bhattacharya, Anirban and Dunson, David B},
  journal={Biometrika},
  volume={98},
  number={2},
  pages={291--306},
  year={2011},
  publisher={Oxford University Press}
}

@article{schiavon2022generalized,
  title={Generalized infinite factorization models},
  author={Schiavon, Lorenzo and Canale, Antonio and Dunson, David B},
  journal={Biometrika},
  volume={109},
  number={3},
  pages={817--835},
  year={2022},
  publisher={Oxford University Press}
}

@article{dahl2006model,
  title={Model-based clustering for expression data via a {Dirichlet} process mixture model},
  author={Dahl, David B},
  journal={Bayesian Inference for Gene Expression and Proteomics},
  volume={4},
  pages={201--218},
  year={2006}
}

@article{biernacki2000assessing,
  title={Assessing a mixture model for clustering with the integrated completed likelihood},
  author={Biernacki, Christophe and Celeux, Gilles and Govaert, G{\'e}rard},
  journal={IEEE Transactions on Pattern Analysis and Machine Intelligence},
  volume={22},
  number={7},
  pages={719--725},
  year={2000},
  publisher={IEEE}
}

@article{neal2000markov,
  title={Markov chain sampling methods for {Dirichlet} process mixture models},
  author={Neal, Radford M},
  journal={Journal of Computational and Graphical Statistics},
  volume={9},
  number={2},
  pages={249--265},
  year={2000},
  publisher={Taylor \& Francis}
}

@article{li2010bayesian,
	title={Bayesian variable selection in structured high-dimensional covariate spaces with applications in genomics},
	author={Li, Fan and Zhang, Nancy R},
	journal={Journal of the American Statistical Association},
	volume={105},
	number={491},
	pages={1202--1214},
	year={2010},
	publisher={Taylor \& Francis}
}

@article{stingo2013integrative,
	title={An integrative {B}ayesian modeling approach to imaging genetics},
	author={Stingo, Francesco C and Guindani, Michele and Vannucci, Marina and Calhoun, Vince D},
	journal={Journal of the American Statistical Association},
	volume={108},
	number={503},
	pages={876--891},
	year={2013},
	publisher={Taylor \& Francis}
}

@article{maynard2021transcriptome,
  title={Transcriptome-scale spatial gene expression in the human dorsolateral prefrontal cortex},
  author={Maynard, Kristen R and Collado-Torres, Leonardo and Weber, Lukas M and Uytingco, Cedric and Barry, Brianna K and Williams, Stephen R and Catallini, Joseph L and Tran, Matthew N and Besich, Zachary and Tippani, Madhavi and others},
  journal={Nature Neuroscience},
  volume={24},
  number={3},
  pages={425--436},
  year={2021},
  publisher={Nature Publishing Group US New York}
}

@article{rovckova2016fast,
  title={Fast {Bayesian} factor analysis via automatic rotations to sparsity},
  author={Ro{\v{c}}kov{\'a}, Veronika and George, Edward I},
  journal={Journal of the American Statistical Association},
  volume={111},
  number={516},
  pages={1608--1622},
  year={2016},
  publisher={Taylor \& Francis}
}

@article{pardo2022spatiallibd,
  title={spatial{LIBD}: an {R/B}ioconductor package to visualize spatially-resolved transcriptomics data},
  author={Pardo, Brenda and Spangler, Abby and Weber, Lukas M and Page, Stephanie C and Hicks, Stephanie C and Jaffe, Andrew E and Martinowich, Keri and Maynard, Kristen R and Collado-Torres, Leonardo},
  journal={BMC Genomics},
  volume={23},
  number={1},
  pages={434},
  year={2022},
  publisher={Springer}
}

@article{jiang2024iimpact,
  title={iIMPACT: integrating image and molecular profiles for spatial transcriptomics analysis},
  author={Jiang, Xi and Wang, Shidan and Guo, Lei and Zhu, Bencong and Wen, Zhuoyu and Jia, Liwei and Xu, Lin and Xiao, Guanghua and Li, Qiwei},
  journal={Genome Biology},
  volume={25},
  number={1},
  pages={1--25},
  year={2024},
  publisher={BioMed Central}
}

@article{yuan2024benchmarking,
  title={Benchmarking spatial clustering methods with spatially resolved transcriptomics data},
  author={Yuan, Zhiyuan and Zhao, Fangyuan and Lin, Senlin and Zhao, Yu and Yao, Jianhua and Cui, Yan and Zhang, Xiao-Yong and Zhao, Yi},
  journal={Nature Methods},
  volume={21},
  number={4},
  pages={712--722},
  year={2024},
  publisher={Nature Publishing Group US New York}
}

@article{song2024scdesign3,
  title={sc{D}esign3 generates realistic in silico data for multimodal single-cell and spatial omics},
  author={Song, Dongyuan and Wang, Qingyang and Yan, Guanao and Liu, Tianyang and Sun, Tianyi and Li, Jingyi Jessica},
  journal={Nature Biotechnology},
  volume={42},
  number={2},
  pages={247--252},
  year={2024},
  publisher={Nature Publishing Group US New York}
}

@article{hao2024dictionary,
  title={Dictionary learning for integrative, multimodal and scalable single-cell analysis},
  author={Hao, Yuhan and Stuart, Tim and Kowalski, Madeline H and Choudhary, Saket and Hoffman, Paul and Hartman, Austin and Srivastava, Avi and Molla, Gesmira and Madad, Shaista and Fernandez-Granda, Carlos and others},
  journal={Nature Biotechnology},
  volume={42},
  number={2},
  pages={293--304},
  year={2024},
  publisher={Nature Publishing Group US New York}
}

@article{hubert1985comparing,
  title={Comparing partitions},
  author={Hubert, Lawrence and Arabie, Phipps},
  journal={Journal of Classification},
  volume={2},
  pages={193--218},
  year={1985},
  publisher={Springer}
}

\clearpage

\begin{table}[]\scriptsize
\caption{Simulated data analysis: Adjust Rand index (ARI) of three Gibbs-type priors, Pitman-Yor process (PY), Dirichlet process (DP), and mixture of finite mixtures (MFM), in the six simulation scenarios. MRFC PY, MRFC DP, and MRFC MFM are MRF-constrained PY, DP, and MFM models.} \label{tab:sim1_ari}
\begin{tabular}{lccclccc}
\hline
\multicolumn{1}{c}{\multirow{2}{*}{\textbf{Method}}} & \multicolumn{3}{c}{\textbf{Strong Signal}}                               &  & \multicolumn{3}{c}{\textbf{Weak Signal}}                                 \\ \cline{2-4} \cline{6-8} 
\multicolumn{1}{c}{}                                 & $H_{0} = 3$                  & $H_{0} = 5$                  & $H_{0} = 7$                   &  & $H_{0} = 3$                   & $H_{0} = 5$                   & $H_{0} = 7$                   \\ \cline{1-4} \cline{6-8} 
PY                                                   & 0.526(0.046)          & 0.698(0.067)          & 0.671(0.035)           &  & 0.232(0.057)          & 0.646(0.089)          & 0.061(0.091)          \\
DP                                                   & 0.531(0.043)          & 0.673(0.092)          & 0.675(0.020)          &  & 0.217(0.067)          & 0.644(0.076)          & 0.036(0.070)          \\
MFM                                                  & \textbf{0.561}(0.095) & \textbf{0.681}(0.096) & \textbf{0.686}(0.019) &  & \textbf{0.265}(0.063) & \textbf{0.656}(0.086) & \textbf{0.106}(0.066) \\
MRFC PY                                   & 0.932(0.014)          & \textbf{0.955}(0.027) & 0.953(0.007)          &  & 0.956(0.013)          & 0.948(0.027)          & 0.961(0.024)          \\
MRFC DP                                   & 0.926 (0.028)          & 0.954 (0.028)          & 0.952 (0.008)          &  & 0.969(0.011)          & 0.937(0.029)          & 0.960(0.027)          \\
MRFC MFM                                  & \textbf{0.936}(0.014) & 0.946(0.023)          & \textbf{0.963}(0.008) &  & \textbf{0.973}(0.023)  & \textbf{0.961}(0.024) & \textbf{0.976}(0.201) \\ \hline
\end{tabular} 
\end{table}

\begin{table}[h!]
\caption{Simulated data analysis: Number of cluster estimation ($\hat{H}$) of three Gibbs-type priors, Pitman-Yor process (PY), Dirichlet process (DP), and mixture of finite mixtures (MFM), in the six simulation scenarios. MRFC PY, MRFC DP, and MRFC MFM are MRF-constrained PY, DP, and MFM models.} \label{tab:sim1_H}
\begin{tabular}{lccclccc}
\hline
\multicolumn{1}{c}{\multirow{2}{*}{\textbf{Method}}} & \multicolumn{3}{c}{\textbf{Strong Signal}}                   &  & \multicolumn{3}{c}{\textbf{Weak Signal}}                     \\ \cline{2-4} \cline{6-8} 
\multicolumn{1}{c}{}                                 & $H_{0} = 3$              & $H_{0} = 5$              & $H_{0} = 7$              &  & $H_{0} = 3$              & $H_{0} = 5$              & $H_{0} = 7$              \\ \cline{1-4} \cline{6-8} 
PY                                                   & 3.4 (0.7)          & 5.3 (0.6)          & 7.4 (0.6)          &  & 1.8 (0.9)          & 5.4 (0.6)          & 2.7(2.2)           \\
DP                                                   & 3.1 (0.3)          & 5.1 (0.3)          & 7.2 (0.4)          &  & 1.9 (0.6)          & 5.1 (0.3)          & 2.3 (1.6)          \\
MFM                                                  & \textbf{3.0 (0.2)} & \textbf{5.0 (0.0)} & \textbf{7.0 (0.1)} &  & \textbf{2.1 (0.5)} & \textbf{5.0 (0.2)} & \textbf{2.9 (1.8)} \\
MRFC PY                                              & 3.0 (0.0)          & \textbf{5.0 (0.0)} & 7.0 (0.0)          &  & 3.2 (0.7)          & 4.8 (0.4)          & 6.8 (0.3)          \\
MRFC DP                                              & 3.0 (0.0)          & 5.0 (0.0)          & 7.0 (0.0)          &  & 3.1 (0.2)          & 4.8 (0.3)          & 6.8 (0.4)          \\
MRFC MFM                                             & \textbf{3.0 (0.0)} & \textbf{5.0 (0.0)} & \textbf{7.0 (0.0)} &  & \textbf{3.0 (0.0)} & \textbf{4.9 (0.3)} & \textbf{6.9 (0.3)} \\ \hline
\end{tabular}
\end{table}

\begin{table}[] \footnotesize
\centering
\caption[Adjust Rand indices of DLPFC samples]{DLPFC and STARmap data analysis: Adjust Rand indices (ARIs) of BNPMFA and competing spatial clustering methods when applied on $8$ DLPFC tissue samples and $3$ STARmap tissue samples} \label{tab:dlpfc_ari}
~\\
\begin{tabular}{ccclcccc}
\hline
\textbf{ID} & \textbf{BNPMFA} & \textbf{BayesSpace} & \textbf{BASS} & \textbf{SpaGCN} & \textbf{SCMEB} & \textbf{DRSC} & \textbf{Louvain} \\ \hline
151507                        & \textbf{0.523}  & 0.420               & 0.447            & 0.346           & 0.426          & 0.425         & 0.363            \\
151508                        & \textbf{0.504}  & 0.316               & 0.406            & 0.257           & 0.425          & 0.349         & 0.328            \\
151509                        & 0.413           & \textbf{0.498}      & 0.431            & 0.318           & 0.468          & 0.421         & 0.304            \\
151510                        & 0.447           & \textbf{0.476}      & 0.423            & 0.303           & 0.398          & 0.319         & 0.269            \\
151669                        & \textbf{0.571}  & 0.369               & 0.384            & 0.269           & 0.285          & 0.228         & 0.178            \\
151670                        & \textbf{0.627}  & 0.226               & 0.378            & 0.387           & 0.263          & 0.309         & 0.156            \\
151671                        & \textbf{0.631}  & 0.389               & 0.558            & 0.233           & 0.306          & 0.364         & 0.263            \\
151672                        & \textbf{0.677}  & 0.572               & 0.571            & 0.291           & 0.436          & 0.417         & 0.274            \\ \hline
BZ5                        & 0.767  & 0.214               & \textbf{0.880}           & 0.248           & 0.186          & 0.299         & 0.152            \\
BZ9                        & \textbf{0.611}  & 0.195               & 0.564            & 0.277           & 0.209          & 0.309         & 0.185            \\
BZ14                        & \textbf{0.820}           & 0.216      & 0.782            & 0.278           & 0.258          & 0.347         & 0.171            \\ \hline
\end{tabular} 
\end{table}

\end{document}


\maketitle

\section{Proof of Proposition 1}

\begin{proof}
Firstly, we calculate the probability $P(\Psi \mid \mathbf{Y}_{0})$ in Definition 1 under the model specification of the proposition. 
\begin{equation*}
\begin{aligned}
 \quad & \int \prod_{h = 1}^{H} \prod_{i:z_{i}= h} f(\bm{y}_{0i};\theta_{h}) dG_{0}(\theta_{h}) \\
 = \quad & C \int \prod_{h = 1}^{H} \prod_{i:z_{i}= h} \tau^{q/2} \det(\Sigma)^{-(n+1)/2} \exp\left[-\frac{1}{2}(\bm{y}_{0i} - \mu_{h})^{T}\Sigma^{-1} (\bm{y}_{0i} - \mu_{h})\right] \\
 \quad \quad & \exp(-\frac{1}{2} \tau \mu_{h}^{T} \Sigma^{-1} \mu_{h}) \det(\Sigma)^{-\frac{q+1}{2}} d(\mu_{h}) d(\Sigma) \\
 = \quad & C \int \prod_{h = 1}^{H} (\frac{\tau}{\tau + n_{h}})^{q/2} \det(\Sigma)^{-\frac{n + q+1}{2}} \exp(-\frac{1}{2} \sum_{i:z_{i} = h} \bm{y}_{0i}^{T} \Sigma^{-1}\bm{y}_{0i} + \frac{1}{2} \frac{n_{h}^{2}}{n_{h} + \tau} \bar{\bm{y}}_{h}^{T} \Sigma^{-1} \bar{\bm{y}}_{h}) d(\Sigma) \\
 = \quad & C \prod_{h=1}^{H}  (\frac{\tau}{\tau + n_{h}})^{q/2} \mid\sum_{h=1}^{H} (\mathbf{S}_{h} + \frac{n_{h}^{2} + 2n_{h}\tau - n_{h} - \tau }{n_{h} + \tau} \bar{\bm{Y}}_{h}\bar{\bm{Y}}_{h}^{T}) \mid ^{-n/2} \\
 =: \quad & f_{\tau}(\Psi, \mathbf{Y}_{0})
\end{aligned}
\end{equation*}
where $n_{h} = \sum_{i=1}^{n} \mathcal{I}(z_{i} = h)$, $\bar{\bm{y}}_{h} = \frac{1}{n_{k}}\sum_{i:z_{i = h}} \bm{y}_{0i}$, and $\mathbf{S}_{h} = \sum_{i:z_{i} = h} (\bm{y}_{0i} - \bar{\bm{y}}_{h})(\boldsymbol{y}_{0i} - \bar{\bm{y}}_{h})^{T}$. The integration is a function of partition $\Psi$ and latent data $\mathbf{Y}_{0}$

Secondly, for any transformation $\mathbf{M}$, such that $\bm y_{i} = \mathbf{M}\bm{y}_{0i}$, we have
\begin{equation*}
    f_{\tau}(\Psi, \mathbf{Y}) = f_{\tau}(\Psi, \mathbf{Y}_{0}) (\det(\mathbf{M}))^{-\frac{n}{2}}
\end{equation*}
with $\mathbf{Y} = (\bm y_{1}, \ldots, \bm y_{n})$. The probability 
\begin{equation*}
    P(\Psi\mid \mathbf{M}\mathbf{Y}_{0}) = \frac{P(\Psi) \times f_{\tau}(\Psi, \mathbf{Y}_{0})\det(\mathbf{M})^{-\frac{n}{2}}}{\sum_{\Psi^{'} \in \mathcal{C}_{n}} P(\Psi^{'}) \times f_{\tau}(\Psi^{'}, \mathbf{Y}_{0})\det(\mathbf{M})^{-\frac{n}{2}}} = P(\Psi \mid \mathbf{Y}_{0})
\end{equation*}
\end{proof}

\section{Additional simulation studies}

\subsection{Evaluation metrics}
We evaluated the performance of BNPMFA in several aspects: clustering performance, and identification of the number of clusters. For clustering results, we quantified performance via the Adjusted Rand Index (ARI) \citep{hubert1985comparing}, a variant of Rand Index (RI), based on estimation of cluster assignment $\bm{\hat{z}} = (\hat{z}_{1}, \ldots, \hat{z}_{n})^{\top}$. Let $A = \sum_{i > i^{'}}{I}(z_{i} = z_{i^{'}}){I}(\hat{z}_{i} = \hat{z}_{i^{'}})$ be the number of pairs belonging to the same group in the truth and estimation; $B = \sum_{i > i^{'}}{I}(z_{i} = z_{i^{'}}){I}(\hat{z}_{i} \neq \hat{z}_{i^{'}})$ be number of pairs which belongs to the same group in the truth but different groups in the estimation; $C = \sum_{i > i^{'}}{I}(z_{i} \neq z_{i^{'}}){I}(\hat{z}_{i} = \hat{z}_{i^{'}})$ be the number of pairs belonging to different groups in the true partition but assigned to the same group in the estimation; $D = \sum_{i > i^{'}}{I}(z_{i} \neq z_{i^{'}}){I}(\hat{z}_{i} \neq \hat{z}_{i^{'}})$, the number of pairs assigned to different groups in both truth and estimation. Then the ARI is defined as
\begin{equation}
    \text{ARI}=\frac{\left(\begin{array}{l}
n \\
2
\end{array}\right)(A+D)-[(A+B)(A+C)+(C+D)(B+D)]}{\left(\begin{array}{l}
n \\
2
\end{array}\right)^2-[(A+B)(A+C)+(C+D)(B+D)]} .\nonumber
\end{equation}
The larger value of ARI indicates a more accurate clustering result. 

\subsection{Scalability test}\label{scalability}
We conducted a scalability test to investigate the scalability of BNPMFA to handle the increasing number of spots $n$ and genes $p$.
The scalability test was conducted in \texttt{R} with \texttt{Rcpp} package using a single thread of an E5 - 2643 v4 CPU (20 M cache, 3.40 GHz) with 256GB memory. 
In the scalability test, we generated spots on a square lattice with $H_{0} = 3$ clusters. 
The spots are arranged on $m \times m$ square lattice where $m = 20, 30, \ldots, 60$, resulting in total $n = m^{2}$ spots in each case. 
Then we generated gene expression molecular according to the simulated data generation scheme with strong signal parameter settings and repeated the steps to generate $20$ independent replicates for each setting of $m$.

\begin{figure}[!h]
	\begin{center}
		\includegraphics[width = 1\textwidth]{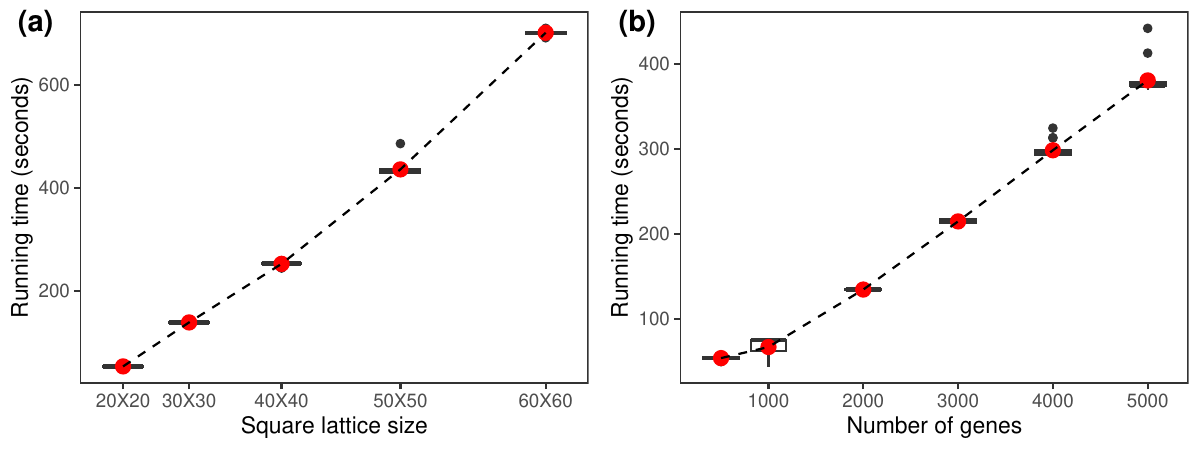}
	\end{center}

\caption{Scalability test. (a) The boxplots of the running time in seconds (per $1,000$ MCMC iterations) in terms of the square pattern size $m^2$ (i.e., number of spots), over $20$ simulated datasets with the number of clusters $H_{0}=3$. (b) The boxplots of the running time in seconds (per $1,000$ MCMC iterations) in terms of the number of genes $p$ over $20$ simulated datasets with the number of clusters $H_{0}=3$. The red points represent the mean computational time of each case.} \label{fig:running_time}
\end{figure}

For each replicate, we applied the proposed model with the same hyperparameters and algorithm settings as described in the simulations of the manuscript. Figure \ref{fig:running_time}(a) demonstrates that the running time of the MCMC algorithm correlates with the square lattice size $m^2$ (i.e., number of spots $n$). We further conducted a linear regression analysis for the running time (per 1,000 iterations) versus $m^2$, the estimated model was as follows: $\text{runtime in seconds} = -46 + 0.2 m^2$, with an adjusted $R^2$ value of $0.990$. Figure \ref{fig:running_time}(b) demonstrates that the running time of the MCMC algorithm correlates with the number of genes $p$. We also conducted a linear regression analysis for the running time (per 1,000 iterations) versus $p$, the estimated model was as follows: $\text{runtime in seconds} = -0.7 + 74 \frac{p}{1000}$, with an adjusted $R^2$ value of $0.992$. 

This result indicates that the running time has a strong linear dependence on the number of spots. 
It implies that, from a computational perspective, BNPMFA is scalable for analyzing SRT data with tens of thousands of spots and genes.

\subsection{Simulated MOB data by scDesign3}
ScDesign3 \citep{song2024scdesign3} is a unified statistical simulator that can generate synthetic spatial omics data. The users can alter parameters to simulate data with realistic features based on the estimated parameters by scDesign3 using real data. We used real SRT data, medial prefrontal cortex STARmap data (BZ5 tissue), as the reference with R function \verb|scdesign3()| to generate the simulated data. To construct the \verb|example_sce| object, we used the count matrix of the gene expression profile and spatial coordinates of BZ5. The augment \verb|cell_type| was set as the manual annotation of spots. Other parameters were the default values in \url{https://songdongyuan1994.github.io/scDesign3/docs/articles/scDesign3-spatial-vignette.html}. 

\begin{figure}[!h]
	\begin{center}
		\includegraphics[width = 1\textwidth]{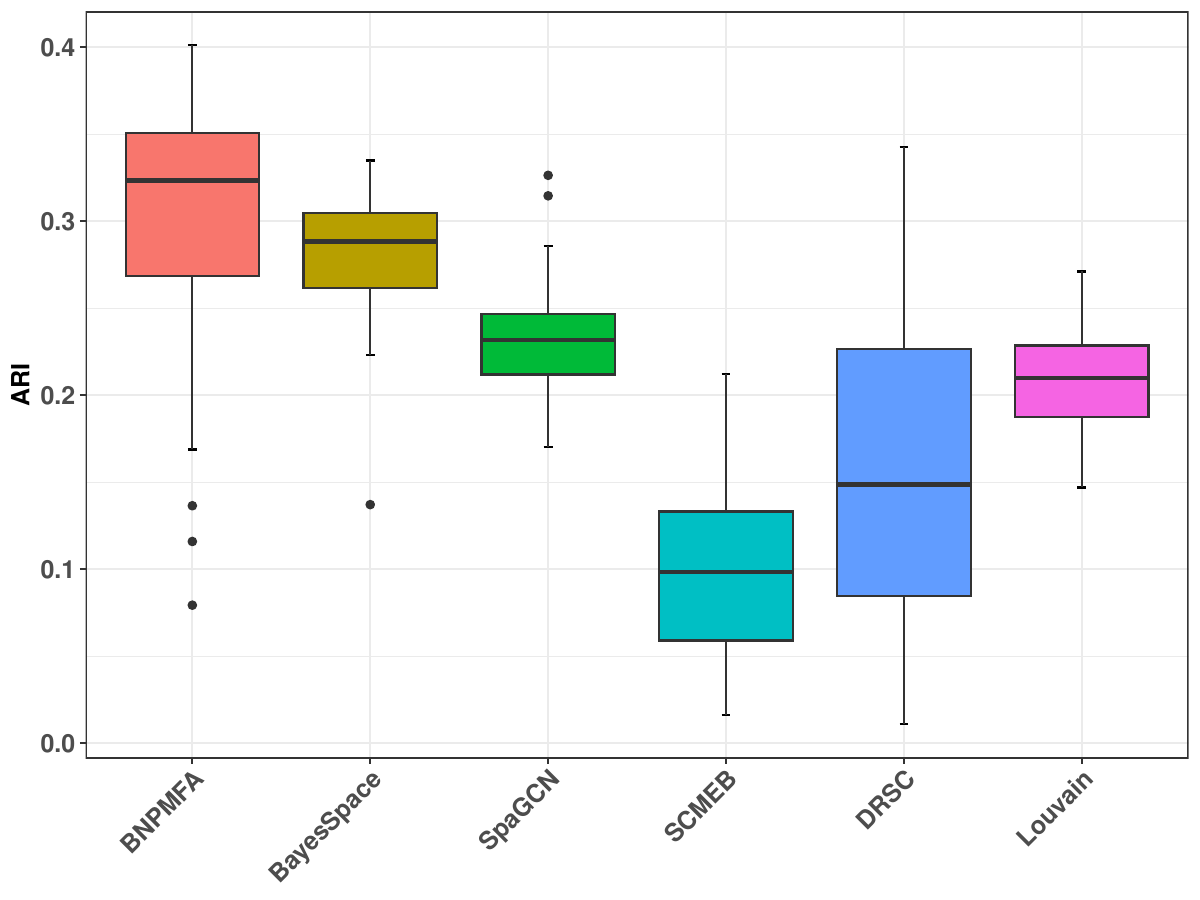}
	\end{center}

\caption{The ARI boxplots for all the methods based on the scDesign3-simulated datasets.} \label{fig:scdesign}
\end{figure}

Using various seed values, we created 50 simulated datasets. Subsequently, we assessed the clustering efficacy of BNPMFA against competing methodologies using these datasets. The results, illustrated in Figure~\ref{fig:scdesign}, demonstrate that BNPMFA surpasses other approaches in terms of clustering performance.

\section{Implementation of the competing methods}
\noindent
\begin{itemize}
    \item BayesSpace: BayesSpace is a Bayesian spatial domain identification method that only leverages
ST data \citep{zhao2021spatial}. The R code of BayesSpace is publicly available on GitHub \url{https://github.com/edward130603/BayesSpace}. The arguments in the main function \verb|spatialCluster()| include preprocessed gene expression matrix with spot coordinates stored in \verb|SingleCellExperiment| object (sce), predetermined region number (q), number of MCMC iterations (nrep), number of iterations in the burn-in period (burn.in), spatial transcriptomic platform (platform), and method for initialization (init.method). In simulation studies and real applications, we let the initialization method be Kmeans and set the predetermined region number as the true value and the number of domains in the annotation, respectively. The other arguments are fixed at the default values.

\item SCMEB: SCMEB is a frequentist approach for spatial domain identification of SRT data \citep{yang2022sc}. The main function is available at R package \verb|SC.MEB|. Firstly, we use the function \verb|find_neighbors2()| to construct an adjacent neighborhood matrix with different platforms (ST or Visium). In the main function \verb|SC.MEB()|, argument (K\_set) is the set of candidates for the number of clusters. In simulation studies and real applications, K\_set is $\{2, \ldots, 10\}$ and other arguments are default settings. Subsequently, we run the model selection function \verb|selectK()| with the same K\_set parameter in the main function to find the optimal model with the BIC criterion.  

\item DRSC: DRSC is a frequentist approach for spatial domain identification of SRT data \citep{liu2022joint}. The main function is available at R package \verb|DR.SC|. Firstly,
we use the function \verb|find_neighbors2()| to construct an adjacent neighborhood matrix with different platforms (ST or Visium). In the main function \verb|DR.SC_fit()|, argument (K) is the set of candidates for the number of clusters. We set K is set $\{2, \ldots, 10\}$. Subsequently, we run the model selection function \verb|selectModel()| with default settings to find the optimal number of clusters. 

\item SpaGCN: SpaGCN is a graph-convolutional-network-based method integrating gene expression, spatial location, and histology to identify spatial domains and spatially variable genes \citep{hu2021spagcn}. The Python code of SpaGCN is publicly available on GitHub \url{https://github.com/jianhuupenn/SpaGCN}. The tutorial in \url{https://github.com/jianhuupenn/SpaGCN/blob/master/tutorial/tutorial.md} provided the details of the implementation where the prespecified number of clusters (n\_clusters) was set as the true number of clusters in simulations and equal to manual annotation in applications. 

\item Louvain: Louvain is an industry-standard method widely utilized in single-cell RNAseq data analysis. It is a graph-based community detection method, integrated into the R package \verb|Seurat| \citep{hao2024dictionary}. According to the grid search method, the resolution parameter was selected from $\{0.1, 0.2, \ldots, 1.5\}$ to ensure the number of clusters equal to true values in simulations and the manual annotations in applications, respectively. The clustering result was obtained under the chosen resolution parameter. 

\item BASS: BASS is a Bayesian spatial segmentation method for SRT data, which achieves contiguous spatial domain identification \citep{li2022bass}. It mainly focuses on single-cell resolution spatial transcriptomics data. The hierarchical model infers the cell types of each spot and the spatial regions induced by these cells. In our simulation settings, BASS cannot find any informative spatial regions, thus we didn't show their results. In the real application of STARmap and DLPFC, BASS provided the spatial domain identification analysis in the tutorial \url{https://zhengli09.github.io/BASS-Analysis/}. Hence, we followed their tutorials and obtained the results in real applications. 

\end{itemize}

\bibliographystyle{agsm}
{\bibliography{supplementary_ref}}

\clearpage

\begin{figure}
\begin{center}
\includegraphics[width = 1\textwidth]{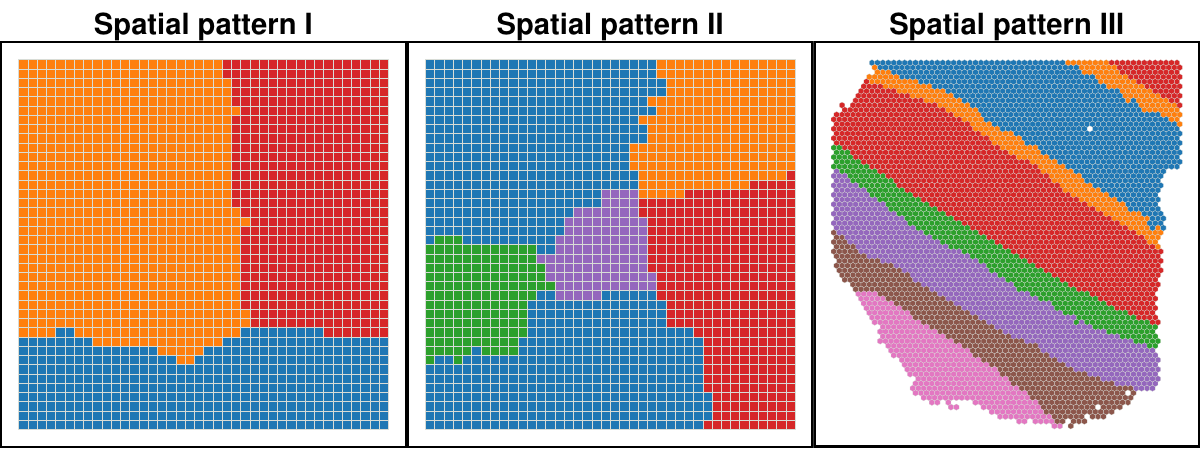}
\end{center}
\caption{The spatial patterns in simulation scenarios: In the spatial pattern I and II, the spots are located in a square lattice. In spatial pattern III, the spots are located in a triangle lattice.  \label{fig:sim_pattern}}
\end{figure}

\begin{figure}[!h]
	\begin{center}
		\includegraphics[width = 1\textwidth]{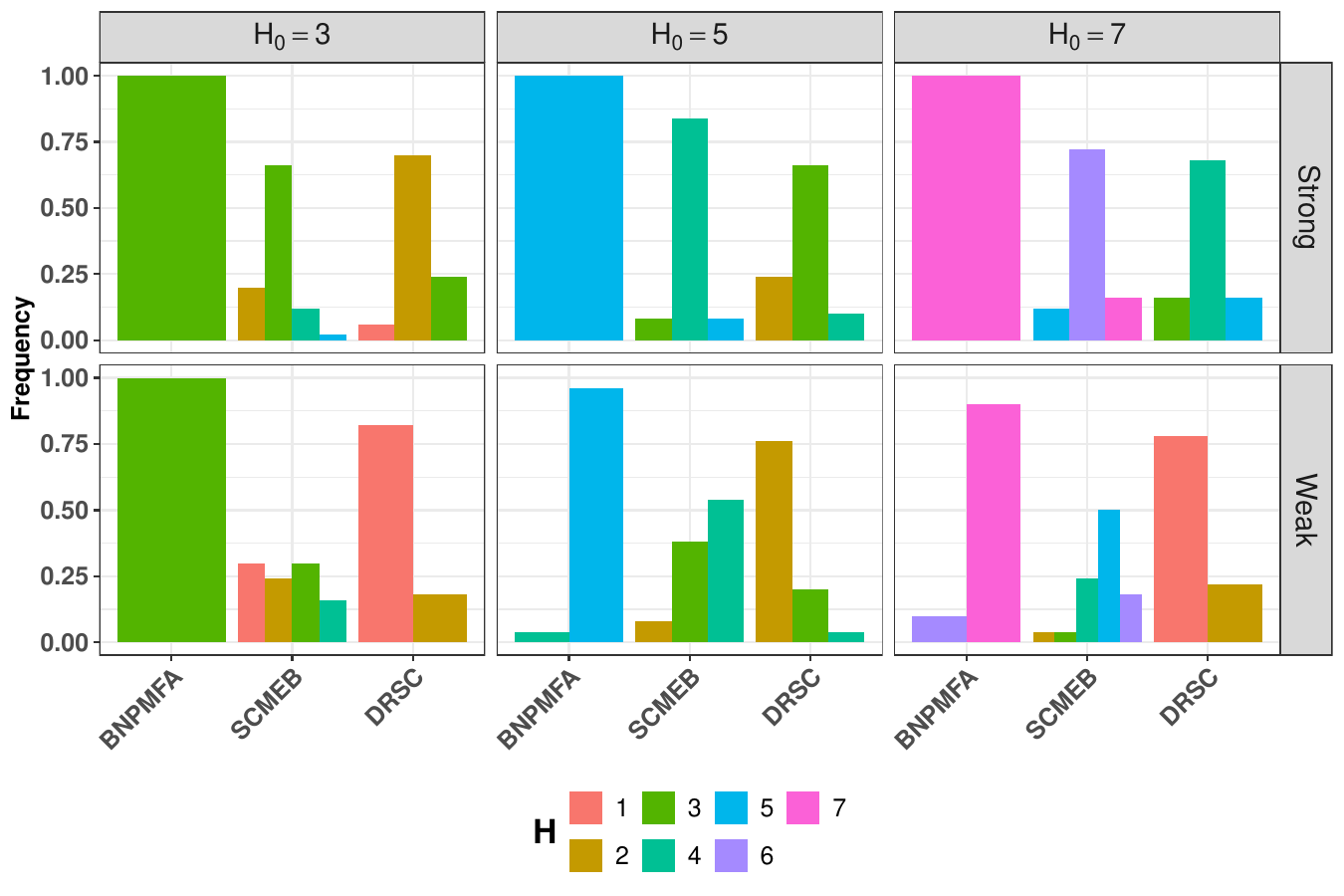}
	\end{center}

\caption{The number of cluster estimation $\hat{H}$ for BNPMFA, SCMEB, and DRSC in $50$ simulation replications of different scenarios. BNPMFA detected true number of clusters in most simulation replications.} 
\end{figure}

\begin{figure}
\begin{center}
\includegraphics[width = 1\textwidth]{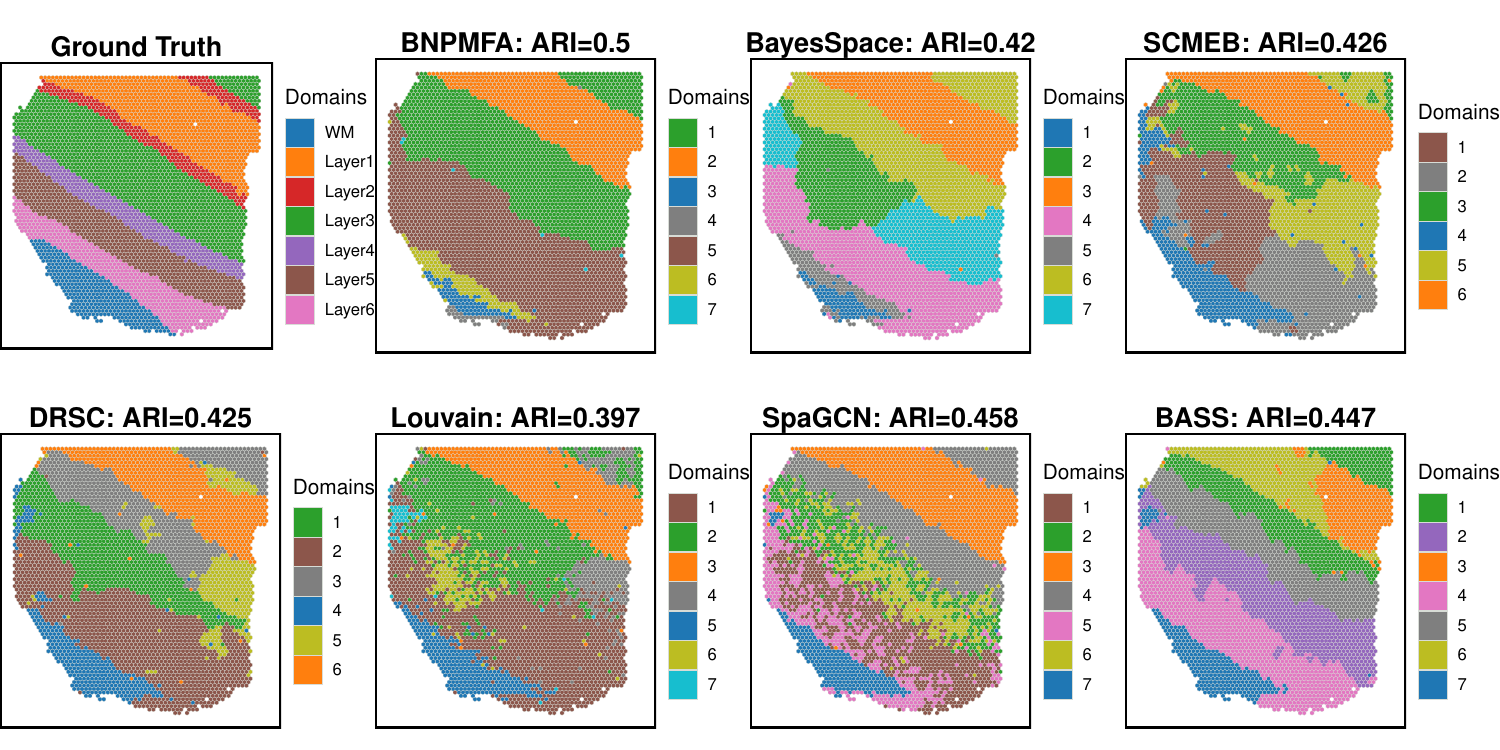}
\end{center}
\caption[Spatial domains of DLPFC 151507 sample]{Spatial domains annotated by pathologists and
identified by BNPMFA and competing methods in DLPFC sample 151507.}
\end{figure}

\begin{figure}
\begin{center}
\includegraphics[width = 1\textwidth]{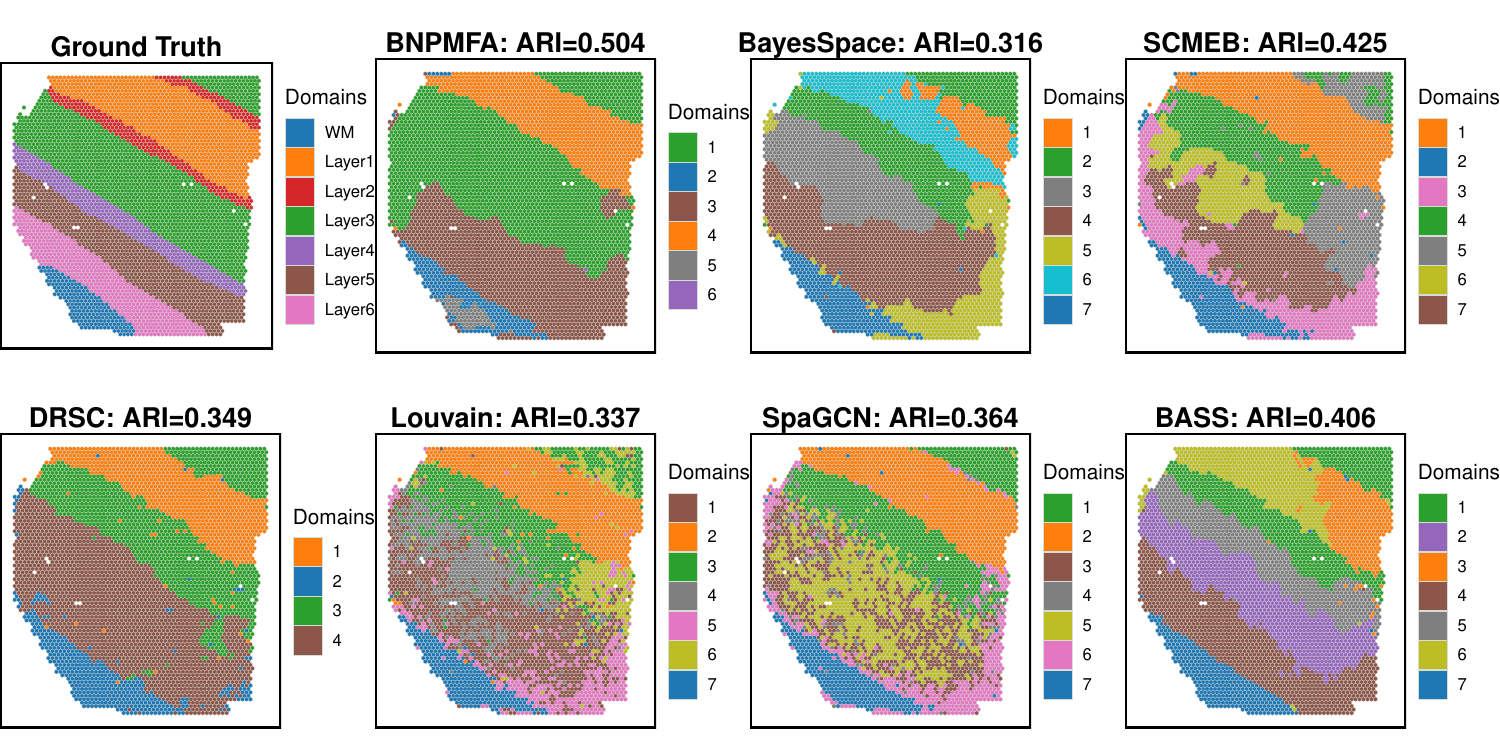}
\end{center}
\caption[Spatial domains of DLPFC 151508 sample]{Spatial domains annotated by pathologists and
identified by BNPMFA and competing methods in DLPFC sample 151508.}
\end{figure}

\begin{figure}
\begin{center}
\includegraphics[width = 1\textwidth]{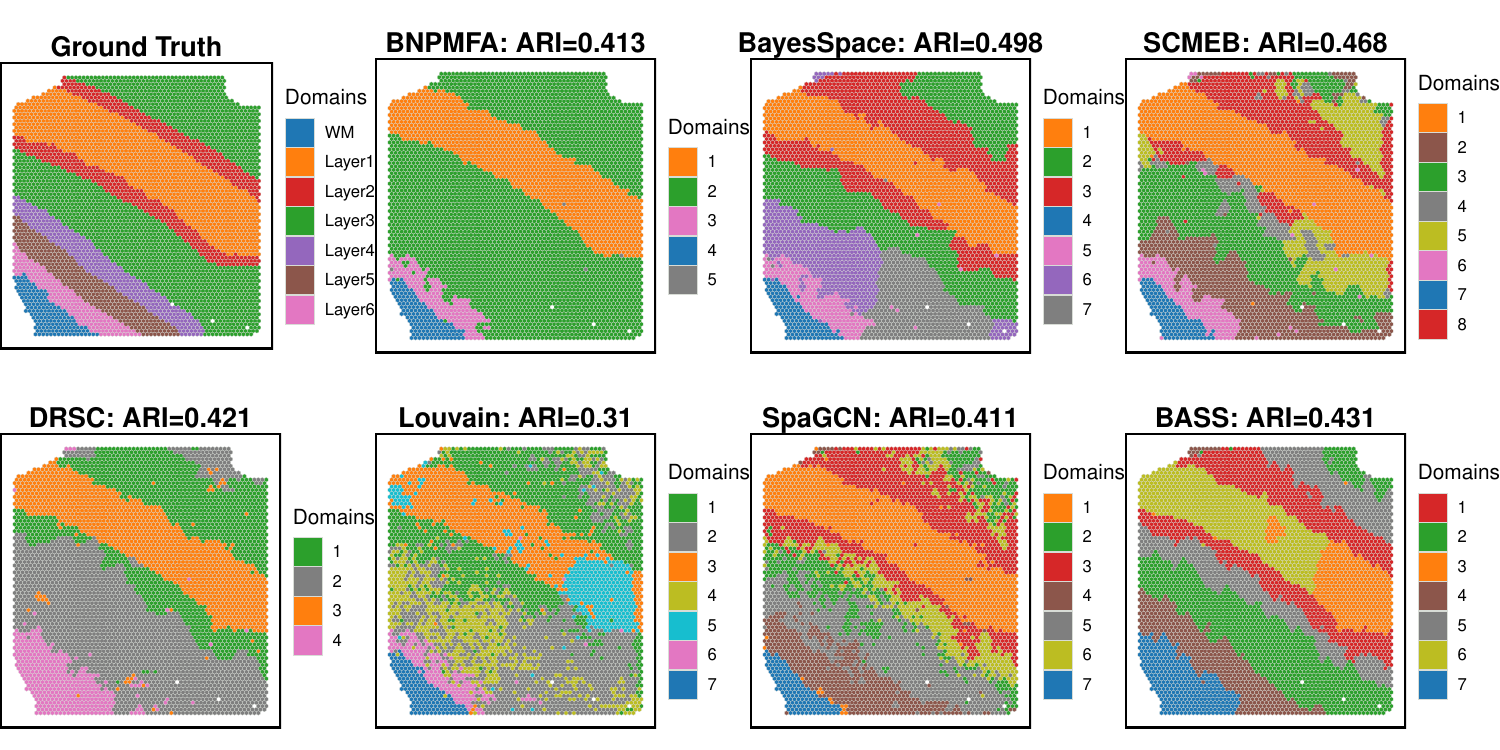}
\end{center}
\caption[Spatial domains of DLPFC 151509 sample]{Spatial domains annotated by pathologists and
identified by BNPMFA and competing methods in DLPFC sample 151509.}
\end{figure}

\begin{figure}
\begin{center}
\includegraphics[width = 1\textwidth]{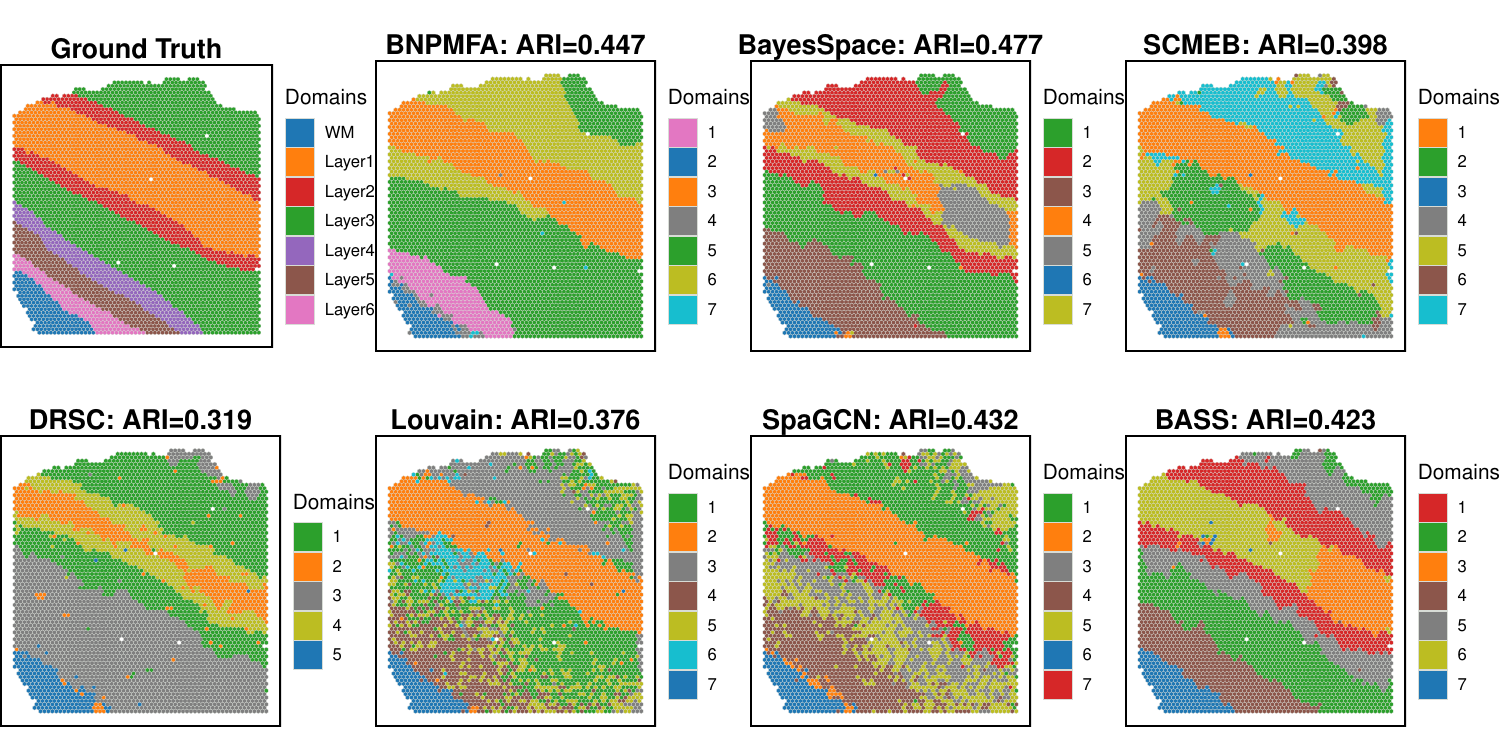}
\end{center}
\caption[Spatial domains of DLPFC 151510 sample]{Spatial domains annotated by pathologists and
identified by BNPMFA and competing methods in DLPFC sample 151510.}
\end{figure}

\begin{figure}
\begin{center}
\includegraphics[width = 1\textwidth]{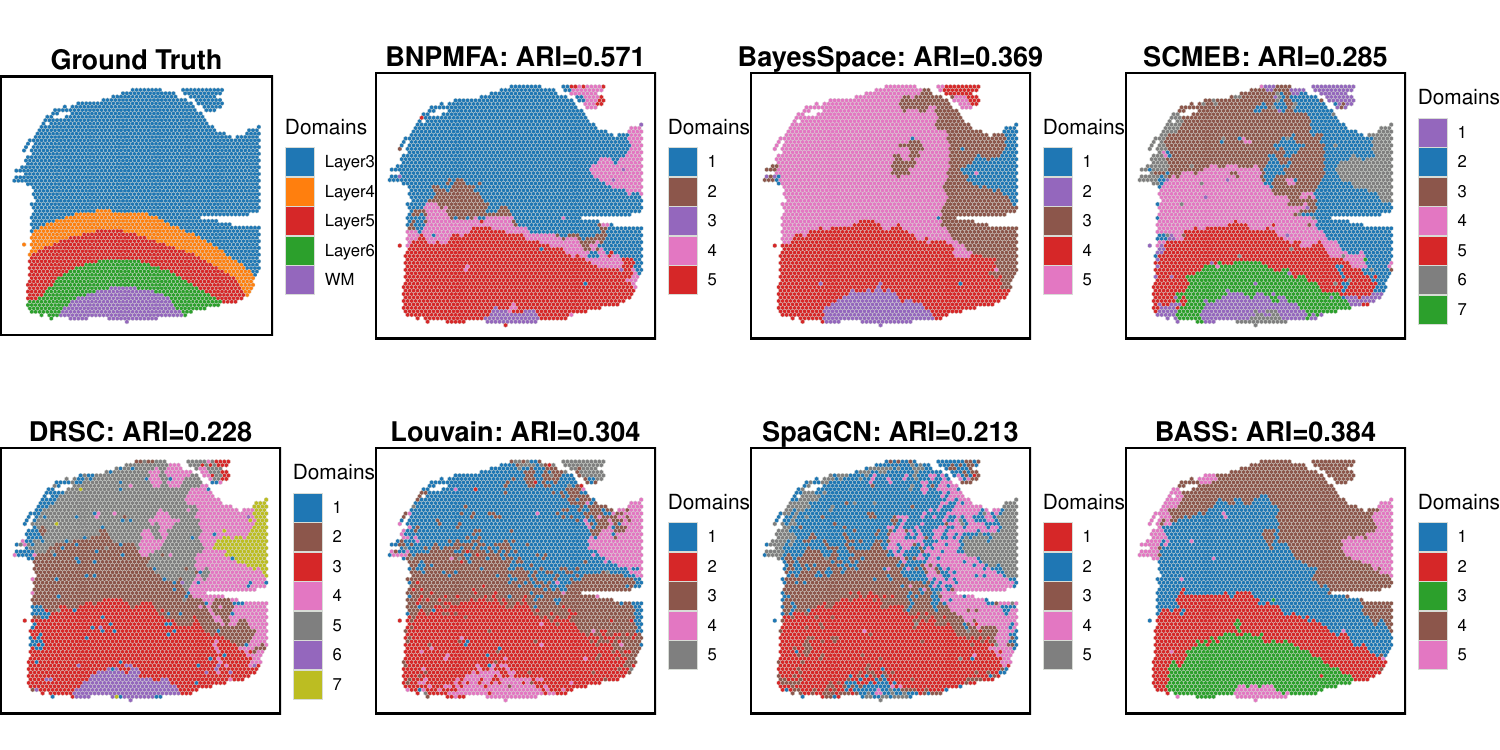}
\end{center}
\caption[Spatial domains of DLPFC 151669 sample]{Spatial domains annotated by pathologists and
identified by BNPMFA and competing methods in DLPFC sample 151669.}
\end{figure}

\begin{figure}
\begin{center}
\includegraphics[width = 1\textwidth]{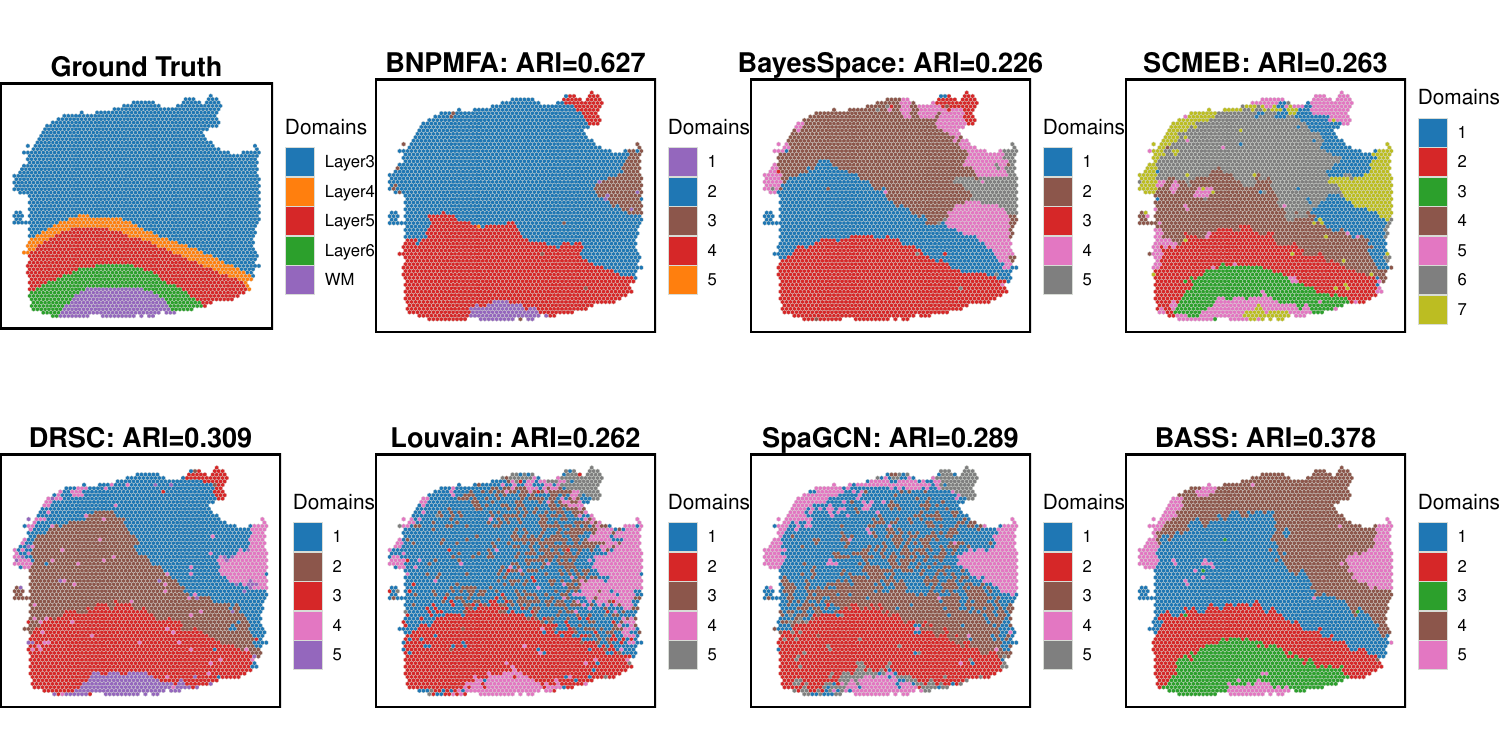}
\end{center}
\caption[Spatial domains of DLPFC 151670 sample]{Spatial domains annotated by pathologists and
identified by BNPMFA and competing methods in DLPFC sample 151670.}
\end{figure}

\begin{figure}
\begin{center}
\includegraphics[width = 1\textwidth]{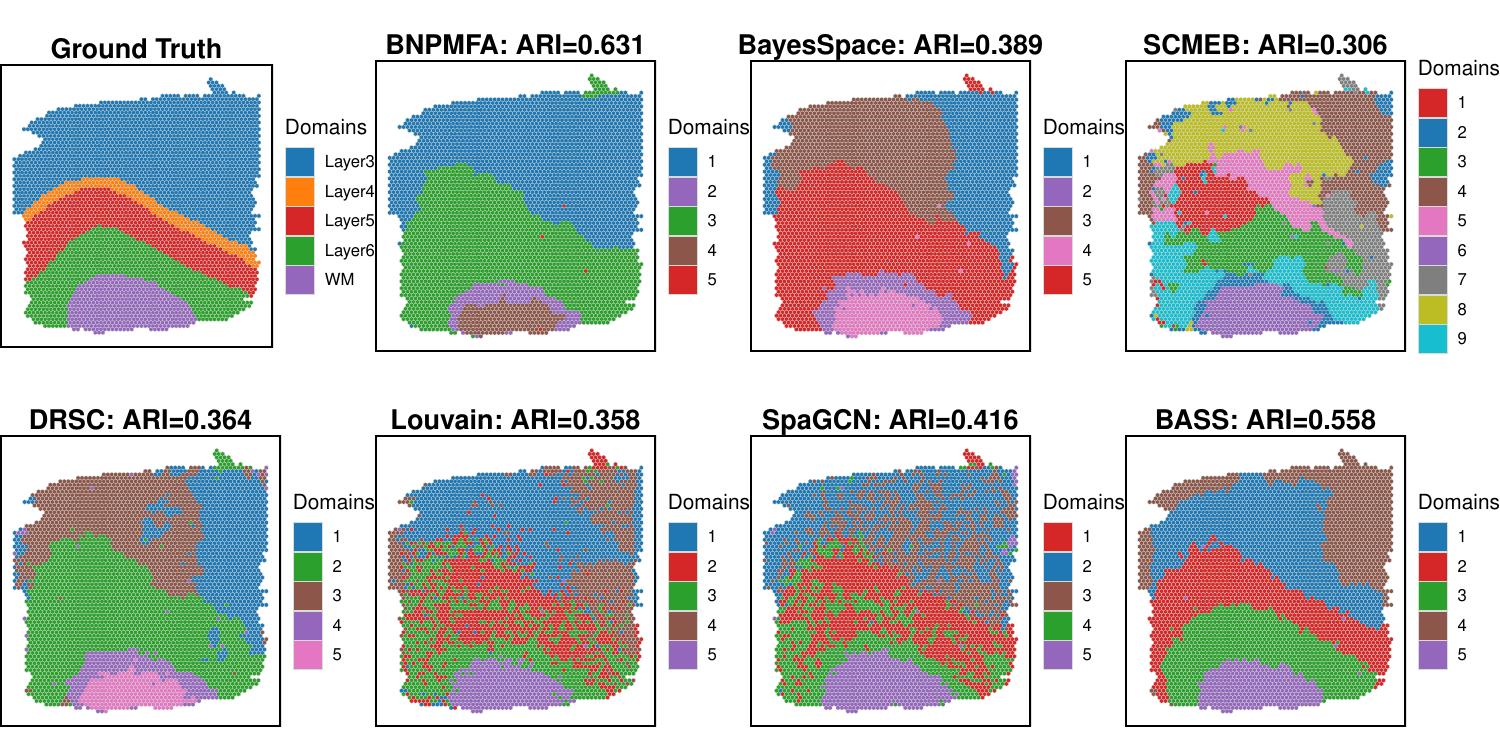}
\end{center}
\caption[Spatial domains of DLPFC 151671 sample]{Spatial domains annotated by pathologists and
identified by BNPMFA and competing methods in DLPFC sample 151671.}
\end{figure}

\begin{figure}
\begin{center}
\includegraphics[width = 1\textwidth]{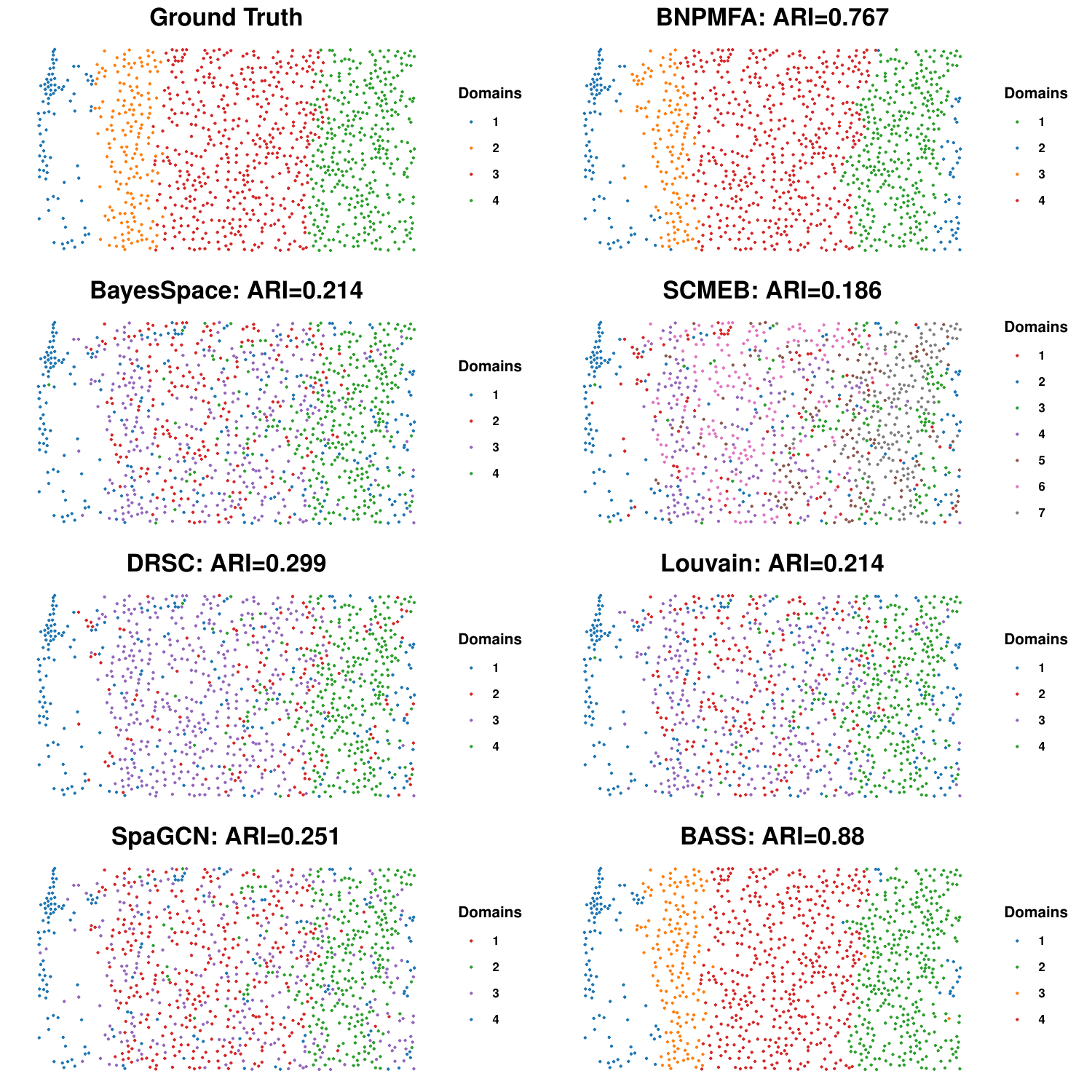}
\end{center}
\caption[Spatial domains of STARmap BZ5 sample]{Spatial domains annotated by pathologists and
identified by BNPMFA and competing methods in STARmap sample BZ5.}
\end{figure}

\begin{figure}
\begin{center}
\includegraphics[width = 1\textwidth]{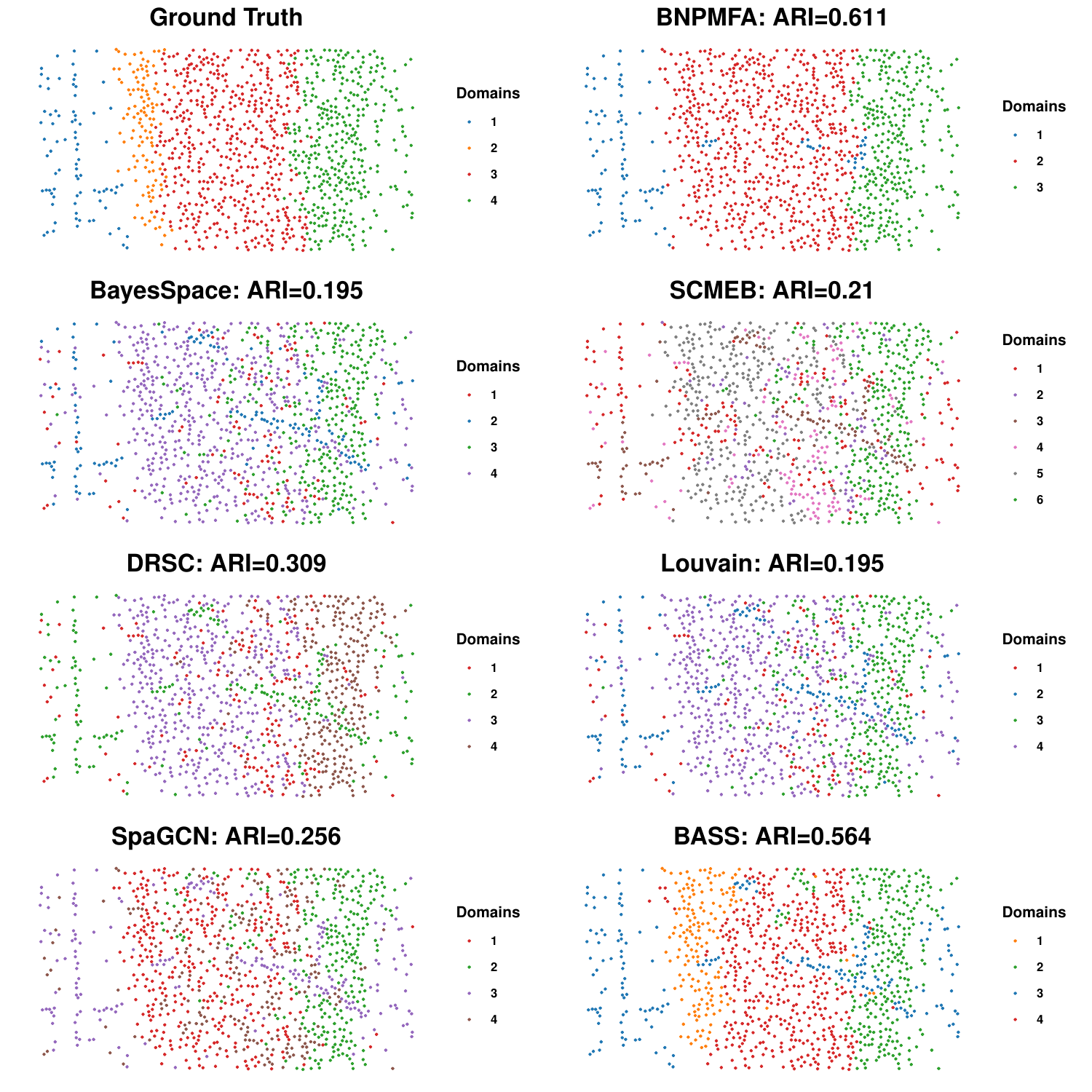}
\end{center}
\caption[Spatial domains of STARmap BZ9 sample]{Spatial domains annotated by pathologists and
identified by BNPMFA and competing methods in STARmap sample BZ9.}
\end{figure}

\clearpage
\begin{table}[]
\centering
\caption{The parameter $\{\mu_{h}\}_{h \geq 1}$ and $\Sigma$ settings in Gaussian mixture component in different simulation signals.} \label{tab:sim_settings}
~\\
\begin{tabular}{lclcl}
\hline 
\multicolumn{1}{c}{\textbf{}} & \multicolumn{2}{c}{\textbf{Strong Signal}}            & \multicolumn{2}{c}{\textbf{Weak Signal}}              \\ \hline
$\mu_{1}$                     & \multicolumn{2}{c}{(5, 0, 0, 0, 0, 0, 0, 0, 0, 0, 0)} & \multicolumn{2}{c}{(3, 0, 0, 0, 0, 0, 0, 0, 0, 0, 0)} \\
$\mu_{2}$                     & \multicolumn{2}{c}{(0, 5, 0, 0, 0, 0, 0, 0, 0, 0, 0)} & \multicolumn{2}{c}{(0, 3, 0, 0, 0, 0, 0, 0, 0, 0, 0)} \\
$\mu_{3}$                     & \multicolumn{2}{c}{(0, 0, 5, 0, 0, 0, 0, 0, 0, 0, 0)} & \multicolumn{2}{c}{(0, 0, 3, 0, 0, 0, 0, 0, 0, 0, 0)} \\
$\mu_{4}$                     & \multicolumn{2}{c}{(0, 0, 0, 5, 0, 0, 0, 0, 0, 0, 0)}                         & \multicolumn{2}{c}{(0, 0, 0, 3, 0, 0, 0, 0, 0, 0, 0)}                         \\
$\mu_{5}$                     & \multicolumn{2}{c}{(0, 0, 0, 0, 5, 0, 0, 0, 0, 0, 0)}                         & \multicolumn{2}{c}{(0, 0, 0, 0, 3, 0, 0, 0, 0, 0, 0)}                         \\
$\mu_{6}$                     & \multicolumn{2}{c}{(0, 0, 0, 0, 0, 5, 0, 0, 0, 0, 0)}                         & \multicolumn{2}{c}{(0, 0, 0, 0, 0, 3, 0, 0, 0, 0, 0)}                         \\
$\mu_{7}$                     & \multicolumn{2}{c}{(0, 0, 0, 0, 0, 0, 5, 0, 0, 0, 0)}                         & \multicolumn{2}{c}{(0, 0, 0, 0, 0, 0, 3, 0, 0, 0, 0)}                         \\ \hline
$\Sigma$                      & \multicolumn{2}{c}{$8 * \mathbb{I}_{10}$}             & \multicolumn{2}{c}{$6 * \mathbb{I}_{10}$}             \\ \hline
\end{tabular}
\end{table}

\begin{table}[h!]
\caption{Simulated data analysis: hyperparameter $d$ tuning in MRF prior for three MRF-constrained Gibbs-type models, MRF-constrained Pitman-Yor process (MRFC PY), MRF-constrained Dirichlet process (MRFC DP), and MRF-constrained mixture of finite mixtures (MRFC MFM). Hyperparameter $d$ is selected from $\{0, 0.5, 1, 1.5, 2, 2.5, 3, 3.5\}$. The values are mean values based on $50$ replications with standard deviation in parentheses.} \label{tab:sim1_H}
~\\
\begin{tabular}{lccclccc}
\hline
\multicolumn{1}{c}{\multirow{2}{*}{\textbf{Method}}} & \multicolumn{3}{c}{\textbf{Strong Signal}}                   &  & \multicolumn{3}{c}{\textbf{Weak Signal}}                     \\ \cline{2-4} \cline{6-8} 
\multicolumn{1}{c}{}                                 & $H_{0} = 3$              & $H_{0} = 5$              & $H_{0} = 7$              &  & $H_{0} = 3$              & $H_{0} = 5$              & $H_{0} = 7$              \\ \cline{1-4} \cline{6-8} 
MRFC PY                                              & 0.72 (0.25)          & 0.93 (0.36) & 0.68 (0.21)          &  & 1.87 (0.92)          & 1.54 (0.76)          & 2.08 (0.65)          \\
MRFC DP                                              & 0.68 (0.15)          & 0.85 (0.22)          & 0.78 (0.21)          &  & 1.98 (0.86)          & 1.51 (0.77)          & 2.08 (0.63)          \\
MRFC MFM                                             & 0.66 (0.22) & 0.80 (0.21) & 0.63 (0.11) &  & 1.60 (0.82) & 1.47 (0.43) & 2.03 (0.69) \\ \hline
\end{tabular}
\end{table}

\begin{table}[]
\centering
\caption{The number of cluster estimation $\hat{H}$ of BNPSpace, SCMEB, and DRSC when applied on $8$ DLPFC samples and  $3$ STARmap tissue samples. $H_{0}$ is the number of clusters in the manual annotation. } \label{tab:app_num_est}
~\\
\begin{tabular}{lcccc}
\hline
\multicolumn{1}{c}{\textbf{ID}} & \textbf{$H_{0}$} & \textbf{BNPMFA} & \textbf{SCMEB} & \textbf{DRSC} \\ \hline
151507                          & 7                & 7               & 6               & 6              \\
151508                          & 7                & 6               & 7               & 4              \\
151509                          & 7                & 5               & 8               & 4              \\
151510                          & 7                & 7               & 7               & 5              \\
151669                          & 5                & 5               & 7               & 7              \\
151670                          & 5                & 5               & 7               & 5              \\
151671                          & 5                & 5               & 9               & 5              \\
{151672}                          & 5              & 5               & 6               & 5              \\ \hline
\multicolumn{1}{c}{BZ5}         & 4                & 4               & 6               & 4              \\
\multicolumn{1}{c}{BZ9}         & 4                & 3               & 6               & 4              \\
\multicolumn{1}{c}{BZ14}        & 4                & 4               & 7               & 5              \\ \hline
\end{tabular}
\end{table}